\documentclass[12pt]{article}
\input{epsf}

\advance\voffset by -1.5cm
\advance\hoffset by -1.3cm
\def\be{\begin{equation}}
\def\ee{\end{equation}}

\def\Zop{\bbbz}

\def\Nop{\bbbn}
\def\Nu{{\cal N}}
\def\N{{\cal N}}

\def\H{{\cal H}}
\def\A{{\cal A}}
\def\R{{\cal R}}
\def\vac{\Omega}
\def\bmu{{\mathord{\hbox{\boldmath$\mu$}}}}
\def\bnu{{\mathord{\hbox{\boldmath$\nu$}}}}
\def\bpsi{\mathord{\hbox{\boldmath$\psi$}}}
\def\bphi{\mathord{\hbox{\boldmath$\phi$}}}

\def\bxi{\mathord{\hbox{\boldmath$\xi$}}}
\def\bnu{\mathord{\hbox{\boldmath$\nu$}}}
\def\bbbone {{\mathchoice {\rm 1\mskip-4mu l} {\rm 1\mskip-4mu l}
{\rm 1\mskip-4.5mu l} {\rm 1\mskip-5mu l}}}
\def\bbbc{{\mathchoice {\setbox0=\hbox{$\displaystyle\rm C$}\hbox{\hbox
to0pt{\kern0.4\wd0\vrule height0.9\ht0\hss}\box0}}
{\setbox0=\hbox{$\textstyle\rm C$}\hbox{\hbox
to0pt{\kern0.4\wd0\vrule height0.9\ht0\hss}\box0}}
{\setbox0=\hbox{$\scriptstyle\rm C$}\hbox{\hbox
to0pt{\kern0.4\wd0\vrule height0.9\ht0\hss}\box0}}
{\setbox0=\hbox{$\scriptscriptstyle\rm C$}\hbox{\hbox
to0pt{\kern0.4\wd0\vrule height0.9\ht0\hss}\box0}}}}
\def\half {\frac{1}{2}}
\def\bbbz {{\sf Z\!\!Z}}

\def\bbbn {{\rm I\!N}}

\def\e{\mathrm{e}}
\def\bomega{{\mathord{\hbox{\boldmath$\omega$}}}}
\def\bOmega{{\mathord{\hbox{\boldmath$\Omega$}}}}
\def\bphi{{\mathord{\hbox{\boldmath$\phi$}}}}
\def\bpsi{{\mathord{\hbox{\boldmath$\psi$}}}}
\def\brho{{\mathord{\hbox{\boldmath$\rho$}}}}
\def\bxi{{\mathord{\hbox{\boldmath$\xi$}}}}
\def\btheta{{\mathord{\hbox{\boldmath$\theta$}}}}

\def\onth{{1\over 3}}
\def\twth{{2\over 3}}

\def\ie {{\it i.e.}}

\begin{document}
\thispagestyle{empty}
\def\thefootnote{\fnsymbol{footnote}}
\vspace*{-1cm}
\begin{flushright}
hep-th/0111260 \\
KCL-MTH-01-46
\end{flushright}

\vspace{1cm}

\begin{center}

{\huge {\bf An algebraic approach to logarithmic}}
\vspace{0.5cm}

{\huge {\bf conformal field theory}}\footnote{Lectures given at the
School on Logarithmic Conformal Field Theory and Its Applications, IPM
Tehran, September 
2001.}   
\vspace{2.0cm}

{\Large Matthias R. Gaberdiel} 
\footnote{{\tt E-mail: mrg@mth.kcl.ac.uk}} \\
\vspace{0.2cm} 

{Department of Mathematics \\ 
King's College London  \\
Strand \\
London WC2R 2LS, U.\ K.\ }
\vspace{2cm}

November 2001
\vspace{2.0cm}

{\bf Abstract}
\end{center}

{\leftskip=0.5truecm
\rightskip=0.5truecm

A comprehensive introduction to logarithmic conformal field theory,
using an algebraic point of view, is given. A number of examples are 
explained in detail, including the $c=-2$ triplet theory and the
$k=-4/3$ affine $su(2)$ theory. We also give some brief introduction
to the work of Zhu.  

}

\newpage

\tableofcontents

\newpage

\setcounter{footnote}{0}
\def\thefootnote{\arabic{footnote}}

\section{Introduction}
\renewcommand{\theequation}{1.\arabic{equation}}
\setcounter{equation}{0}

In the last few years conformal field theories whose correlation
functions have logarithmic branch cuts have attracted some
attention. The interest in these {\it logarithmic} conformal field
theories has been motivated from two different points of view. First 
of all, models with this property seem to play an important role for
the description of certain statistical models, in particular in the
theory of (multi)critical polymers \cite{Sal92,Flohr95,Kau95},
percolation \cite{Watts96}, two-dimensional turbulence
\cite{RR95,Flohr96b,RR96}, the quantum Hall effect
\cite{GurFloNay97,Ino98}, the sandpile model \cite{MatRue} and
disordered systems \cite{CKT95,CKT96,MSer96,CTT98,gll,ludwig}. 
Secondly, this new class of conformal field theories is of conceptual
interest since it is likely to shed light on the structure of more
general conformal field theories beyond the familiar rational
models. In particular, there exists a family of `rational' logarithmic 
conformal field theories, the simplest of which is the triplet algebra
at $c=-2$ \cite{GKau96b,GKau98}, that define in some sense rational
theories, but whose structure differs quite significantly from what
one expects based on the experience with, say, affine theories at
integer level. For example, as we shall explain in detail, the
Verlinde formula \cite{verlinde} does not hold for these examples, and 
neither do the polynomial relations of  Moore \& Seiberg \cite{ms}. If
we are interested in understanding the structure of general (rational)
conformal field theories, it is therefore necessary to get to grips
with these logarithmic conformal field theories.

By now, quite a number of logarithmic models have been analysed. They
include the WZW model on the supergroup $GL(1,1)$ \cite{RSal92}, the
$c=-2$ model \cite{Gur93,GKau96a}, gravitationally dressed conformal
field  theories \cite{BKog95}, and WZW models at $k=0$ 
\cite{CKT95,CKT96,Nichols,Nichols1} and at fractional level
\cite{Gab01}. Singular vectors of some Virasoro 
models have been constructed in \cite{Flohr98} (see also \cite{FL}), 
correlation functions have been calculated in \cite{SR,RMK,KAR,Flohr01}
(see also \cite{RT}), and more structural properties of the
representation theory have been analysed in \cite{Roh}. Recently it
has also been attempted to construct boundary states for a logarithmic
conformal field theory \cite{KoW,Ish,KaW} (see also \cite{Ishi,Kawai}).
\smallskip

In these lecture notes we want to give a comprehensive introduction to
logarithmic conformal field theory. Our approach will be mostly
algebraic in that we shall attempt to describe the characteristic
features in terms of the representation theory of the Virasoro
algebra. We shall discuss in detail the rational logarithmic model
\cite{GKau96b} and its corresponding local theory \cite{GKau98}. We
shall also explain the essential features of the logarithmic conformal
field theory corresponding to the WZW model of $su(2)$ at fractional
level \cite{Gab01}. We shall assume a basic familiarity with the
concepts of conformal field theory\footnote{Introductions to conformal
field theory can be found in \cite{Ginspargrev,DMS}; see also
\cite{Gabrev} for a treatment in the spirit of these lectures.}, but
we shall also, every now and then, pause to explain some more general
facts about conformal field theory that may not be well known; in
particular, we shall give a brief introduction to the work of Zhu
\cite{Zhu}. 
\medskip

These lectures are organised as follows: in section~2 we begin with the 
simplest example of a logarithmic conformal field theory, the Virasoro
model at $c=-2$ \cite{Gur93}. We explain how logarithmic branch cuts
arise by solving the differential equations that determine a certain
four-point function. We then show how this result can be recovered
from an algebraic point of view in terms of an indecomposable
representation of the Virasoro algebra. Section~3 deals with the
(chiral) triplet model \cite{GKau96b}. We begin by describing the
various highest weight representations of this theory (using ideas of
Zhu), and we explain how to determine the fusion products of these 
representations. We also comment on the characters of these
representations and their modular properties. In section~4 we then
discuss how to construct the corresponding local theory
\cite{GKau98}. As we shall see, the locality constraint determines the
theory (and its structure constants) uniquely, and the corresponding
partition function is then automatically modular invariant. Section~5
deals with another example of a (chiral) logarithmic conformal field
theory, the WZW model corresponding to $su(2)$ at fractional level
\cite{Gab01}. We close with some conclusions in section~6. We have
included an appendix in which details of a key calculation in the
construction of Zhu's algebra are given. 

\newpage

\section{The baby example}
\renewcommand{\theequation}{2.\arabic{equation}}
\setcounter{equation}{0}

In this section we want to describe the simplest example of a
logarithmic representation for the $c=-2$ Virasoro theory. We shall do
the analysis in two different ways, first by solving the differential
equation that characterises a certain correlation function, and then
by analysing the corresponding fusion product algebraically. This will
demonstrate how these two different points of view are related to one
another. The description of the analytic approach is due to Gurarie
\cite{Gur93} who was the first to study conformal correlators with
logarithmic branch cuts systematically. The algebraic approach was
first applied to this problem in \cite{GKau96a}. 

\subsection{The analytic approach}

We are interested in determining the `fusion product' of the $c=-2$
representation with highest weight $\mu$ for which 
\be\label{muh}
L_n\, \mu =0 \quad \hbox{for $n>0$,}  \qquad 
L_0 \,\mu = - {1\over 8} \mu
\,. 
\ee
To this end we want to determine the four-point function 
\be\label{4pt}
\langle \mu(z_1) \mu(z_2) \mu(z_3) \mu(z_4) \rangle \,.
\ee
Using the M\"obius symmetry (\ie\ the kinematical constraints that
have been discussed in the lectures by Rahimi Tabar \cite{RT}) 
we can write this function as
\be
\langle \mu(z_1) \mu(z_2) \mu(z_3) \mu(z_4) \rangle =
(z_1-z_3)^{1 \over 4} (z_2-z_4)^{1\over 4} 
\left[ x (1-x) \right]^{1\over 4} F(x) \,,
\ee
where $F$ is a function that is yet to be determined (by dynamical
constraints), and $x$ is the anharmonic ratio,
\be
x= {(z_1-z_2) (z_3-z_4) \over (z_1-z_3) (z_2-z_4)} \,.
\ee
In order to obtain a constraint for the function $F$ we use the fact
that the representation generated from $\mu$ contains the null-vector,
\be\label{null}
\N = \left( L_{-2} - 2 L_{-1}^2 \right) \mu \,.
\ee
\medskip

\noindent {\it Exercise:} Check that $\N$ is a null-vector, \ie\ that 
$L_1\, \N = L_2\, \N = 0$, using the commutation relations of the
Virasoro algebra
\be
[L_m,L_n] = (m-n) L_{m+n} + {c\over 12} m (m^2-1) \delta_{m,-n} 
\ee
with $c=-2$, as well as (\ref{muh}). 
\medskip

\noindent Next we use the fact that the null-vector must vanish in all
correlation functions, and therefore that in particular,
\be\label{corr1}
0 = \langle \mu(z_1) \mu(z_2) \mu(z_3) 
\left[(L_{-2} - 2 L_{-1}^2)\mu\right](z_4) \rangle
\,.
\ee
As we have heard in the lectures of Rahimi Tabar and Flohr
\cite{RT,FL},  
\be\label{deri1}
\left[L_{-1} \mu\right](z) = {d \over dz} \mu(z) \,,
\ee
and thus the term with $L_{-1}^2$ becomes 
${\partial^2 \over \partial z_4^2}$. In order to evaluate the term
with $L_{-2}$ further, we observe that 
\be\label{lm2}
L_{-2}\mu(z_4) = \oint_{z_4} dz {1\over (z-z_4)} T(z) \, \mu(z_4)\,,
\ee
where $T(z)$ is the stress-energy tensor with mode expansion
\be
T(z) = \sum_{n\in\Zop} L_n z^{-n-2} \,.
\ee
We can thus rewrite $L_{-2}$ in (\ref{corr1}) using (\ref{lm2}). The
contour can then be `pulled off' from $z_4$ so that the integral can
be evaluated in terms of contour integrals around each of $z_1,z_2$
and $z_3$. Using the operator product expansion of $T(z)$ with
$\mu(w)$, 
\be
T(z) \mu(w) = -{1\over 8} {1 \over (z-w)^2} + {1\over (z-w)} 
{d\over dw} \mu(w) + O(1) \,,
\ee
this can thus be written in terms of derivatives acting on $z_i$ 
with $i=1,2,3$. After a small calculation one then arrives at the 
differential equation for $F$, 
\be\label{diff}
x(1-x) F''(x) + (1-2x) F'(x) - {1\over 4} F(x) = 0 \,.
\ee
The usual technique of constructing a solution to this differential
equation is to make the ansatz
\be\label{ansatz}
F(x) = G(x) = x^s ( a_0 + a_1 x + a_2 x^2 + \ldots) \,, 
\qquad a_0\ne 0 \,,
\ee
and then to determine the coefficients $a_i$ recursively. In order for
this to be possible, we first have to solve the `indicial equation',
\ie\ we have to solve the term proportional to $x^{s-1}$ in
(\ref{diff}). (This will determine the value of $s$.) In our case, the
indicial equation is simply
\be
s(s-1) + s = s^2 = 0 \,.
\ee
Thus the two solutions (or roots) for $s$ coincide. It thus follows,
that only one solution to (\ref{diff}) is of the form (\ref{ansatz}),
while the other is 
\be\label{second}
F(x) = G(x) \log(x) + H(x) \,,
\ee
where $G(x)$ is the solution (\ref{ansatz}), and both $G$ and $H$ are
regular at $x=0$. In our case, the second solution (\ref{second}) is
actually equal to $G(1-x)$, where $G$ is given by
(\ref{ansatz}). Thus the most general solution is  
\be\label{gensol}
F(x) = A G(x) + B G(1-x) \,,
\ee
where $A$ and $B$ are arbitrary constants and we have
\be
G(1-x) = G(x) \log(x) + H(x) \,.
\ee
It thus follows that the four-point function (\ref{4pt}) has
necessarily a {\it logarithmic} singularity somewhere on the Riemann
sphere. Indeed, the solution (\ref{gensol}) has a logarithmic
singularity at $x=0$ unless $B=0$, and a logarithmic singularity at
$x=1$ unless $A=0$.
\bigskip

We can interpret the result for the correlation function in terms of
the operator product expansion of $\mu$ with itself, 
\be\label{opelog}
\mu(z) \mu(0) \sim z^{1\over 4} 
\left( \omega(0) + \Omega(0) \log(z) \right) + \ldots\,.
\ee
Here the two states $\omega$ and $\Omega$ label the two different
solutions of the differential equations (that correspond to the two
constants $A$ and $B$).

\subsection{The algebraic approach}

We shall now explain how these results can be understood in terms of
the representation theory of the Virasoro algebra. From an algebraic
point of view, fusion can be regarded as some sort of tensor product
of representations, that associates to two representations of the
Virasoro algebra (the representations inserted at $z=z_1$ and $z=z_2$)
a product representation.\footnote{We should stress, however, that
this product representation is not simply the tensor product; in fact,
the usual tensor product representation has a central charge that is
the sum of the two central charges, but we are here interested in
obtaining a representation with the same central charge.} In fact, by
considering the contour integral of the stress energy tensor around
both insertion points, a product of highest weight states naturally
defines a representation. Using this as the definition of the product
representation, the action of the Virasoro modes on this `tensor
product' can then be described in terms of a comultiplication formula    
\begin{eqnarray}
\label{m}
\oint_{0} dw \; w^{m+1} \hspace*{-0.5cm} & & 
\left\langle\phi| T(w) \;V(\psi,z_1) \; 
V(\chi,z_2)\;\Omega \right\rangle \nonumber \\
& & = \sum
\left\langle\phi| V(\Delta^{(1)}_{z_1,z_2}\,(L_{m})\,\psi,z_1)\;
V(\Delta^{(2)}_{z_1,z_2}\,(L_{m})\,\chi,z_2)\;\Omega \right\rangle\,,
\end{eqnarray}
where $\phi$ is an arbitrary state inserted at infinity, and we have
written  
\begin{equation}
\Delta_{z_1,z_2}(L_m) = \sum \Delta_{z_1,z_2}^{(1)}(L_m)\otimes 
\Delta_{z_1,z_2}^{(2)}(L_m)\,.
\end{equation}
More explicitly, the comultiplication formula is given by
\cite{Gab93} 
\begin{equation}
\label{vir1}
{\displaystyle \Delta_{z_1,z_2}(L_{n})} = 
{\displaystyle \sum_{m=-1}^{n} \left( \begin{array}{c} n+1 \\ m+1
\end{array} \right)
z_1^{n-m} \left(L_{m} \otimes \bbbone \right) +
\sum_{l=-1}^{n} \left( \begin{array}{c} n+1 \\ l+1
\end{array} \right)
z_2^{n-l} \left(\bbbone \otimes L_{l} \right)\,,} \hspace*{0.4cm} 
\end{equation}
where $n\geq -1$. For $n\leq -2$, there are two different formulae
that correspond to two different expansions of the same meromorphic
function, 
\begin{eqnarray}
{\displaystyle \Delta_{z_1,z_2}(L_{-n})} = &
{\displaystyle \sum_{m=-1}^{\infty} \left( \begin{array}{c} n+m-1 \\ m+1
\end{array} \right) (-1)^{m+1}
z_1^{-(n+m)} \left(L_{m} \otimes \bbbone \right)+ } \hspace*{1.5cm}
\nonumber \\
\label{vir2}
& \hspace*{0.5cm} {\displaystyle
\sum_{l=n}^{\infty} \left( \begin{array}{c} l-2 \\ n-2
\end{array} \right)
(-z_2)^{l-n} \left(\bbbone \otimes L_{-l} \right)},
\end{eqnarray}
and
\begin{eqnarray}
{\displaystyle \widetilde\Delta_{z_1,z_2}(L_{-n})} = &
{\displaystyle
\sum_{l=n}^{\infty} \left( \begin{array}{c} l-2 \\ n-2
\end{array} \right)
(-z_1)^{l-n} \left(L_{-l} \otimes \bbbone\right) + } \hspace*{3cm}
\nonumber \\ 
\label{vir21}
& \hspace*{0.5cm} 
{\displaystyle \sum_{m=-1}^{\infty} \left( \begin{array}{c} n+m-1 \\ m+1
\end{array} \right) (-1)^{m+1}
z_2^{-(n+m)} \left(\bbbone \otimes L_m\right)\,,} 
\end{eqnarray}
where in both formulae $n\geq 2$. The two formulae (\ref{vir2}) and 
(\ref{vir21}) must agree in all correlation functions (since they were
both derived from (\ref{m})); the actual fusion product 
$(\H_1 \otimes \H_2)_{\mathrm{f}}$ of the two representations $\H_1$
and $\H_2$ is therefore the quotient space of the product space where
we quotient out states of the form  
\be\label{fusionquotient}
\left[ \Delta_{z_1,z_2}(L_m) - \widetilde\Delta_{z_1,z_2}(L_m) \right]
(\psi_1\otimes \psi_2) \,,
\ee
where $m\in\Zop$ and $\psi_i\in\H_i$ are arbitrary states. The action
of the Virasoro algebra on $(\H_1 \otimes \H_2)_{\mathrm{f}}$ is then
given by (\ref{vir1}) as well as either (\ref{vir2}) or
(\ref{vir21}). 

In general it is difficult to analyse the whole representation
space. However, we can analyse various quotient spaces fairly easily,
in particular the `highest weight space' \cite{Nahm}
\be\label{highest}
\left( \H_1 \otimes \H_2 \right)^{(0)}_{\mathrm{f}} = 
\left( \H_1 \otimes \H_2 \right)_{\mathrm{f}} / \A_{-} 
\left( \H_1 \otimes \H_2 \right)_{\mathrm{f}} \,,
\ee
where $\A_{-}$ is the algebra of negative modes, \ie\ the algebra
generated by $L_{-n}$ with $n>0$. This quotient space describes the
dual of the highest weight space, \ie\ it consists of those states
that have a non-trivial correlation function with a highest weight
state at infinity. There are also more complicated and bigger quotient
spaces that can be determined and that uncover more structure of the
actual fusion product \cite{GKau96a}; in these lectures we shall
however only work out the space (\ref{highest}) explicitly. 

The analysis of the various quotient spaces is simplified by choosing
$z_1=1$ and $z_2=0$. Furthermore, one can use the fact that for 
$n\geq 2$, 
\be\label{2.26}
\widetilde{\Delta}_{0,-1}(L_{-n}) = \widetilde\Delta_{1,0}
\left(\e^{L_{-1}} L_{-n} \e^{-L_{-1}}\right) \cong
\Delta_{1,0}\left(\e^{L_{-1}} L_{-n} \e^{-L_{-1}}\right) \,,
\ee
where the last identity holds in the quotient space 
$\left( \H_1 \otimes \H_2 \right)_{\mathrm{f}}$ because of
(\ref{fusionquotient}). The right hand side of (\ref{2.26}) consists
of negative modes only; to obtain the quotient space (\ref{highest})
we can therefore divide out any state of the form 
$\widetilde{\Delta}_{0,-1}(L_{-n})\psi$ with $n\geq 2$ (as well as
obviously states of the form $\Delta_{1,0}(L_{-n})$ with $n\geq 1$). 
The relevant comultiplication formulae are then 
\be
\Delta_{1,0}(L_{-n}) = 
\sum_{m=-1}^{\infty} \left( \begin{array}{c} n+m-1 \\ m+1
\end{array} \right) (-1)^{m+1}
\left(L_{m} \otimes \bbbone \right)+ 
\left(\bbbone \otimes L_{-n} \right)\,,
\ee
and
\be
\widetilde\Delta_{0,-1}(L_{-n}) = 
\left(L_{-n} \otimes \bbbone\right) + 
\sum_{m=-1}^{\infty} \left( \begin{array}{c} n+m-1 \\ m+1
\end{array} \right) (-1)^{n+1}
\left(\bbbone \otimes L_m\right)\,,
\ee
where again $n\geq 2$. Let us now apply this analysis to the fusion
product of $\H_{\mu}$ with itself, where $\H_{\mu}$ denotes the
irreducible highest weight representation generated from the highest
weight state $\mu$ satisfying (\ref{muh}). Using the above relations 
repeatedly (together with (\ref{vir1}) with $n=-1$ and the null vector
relation (\ref{null})) it is not difficult to see that the highest
weight space  
\be
\left( \H_{\mu} \otimes \H_{\mu} \right)_{\mathrm{f}}^{(0)}
\ee
can be taken to be spanned by the two vectors
\be\label{basis}
(\mu \otimes \mu)  \qquad \hbox{and}  \qquad 
(L_{-1} \mu \otimes \mu) \,.
\ee
Next, we want to determine the action of $L_0$ on this two-dimensional
highest weight space. Using (\ref{vir1}) (with $z_1=1$, $z_2=0$ and
$n=0$) and dropping the comultiplication symbol $\Delta$ where no
confusion can arise, we obtain 
\begin{eqnarray}\label{L0calc}
L_0 (\mu\otimes \mu) & = & (L_{-1}\mu \otimes \mu) - {1\over 4}
(\mu\otimes \mu) \nonumber \\
L_0 (L_{-1} \mu\otimes \mu) & = & (L_{-1}^2\mu \otimes \mu) 
+ {3\over 4} (L_{-1}\mu\otimes \mu)\,.
\end{eqnarray}
Using the null vector (\ref{null}) we have 
$L_{-1}^2\mu = {1\over 2} L_{-2}\mu$; thus we obtain
\begin{eqnarray}
(L^2_{-1} \mu\otimes \mu) & = & {1\over 2} (L_{-2}\mu \otimes \mu)
\nonumber \\
& \cong & {1\over 2} (L_{-2}\mu \otimes \mu) -
{1\over 2} \widetilde\Delta_{0,-1}(L_{-2})(\mu\otimes \mu)
\nonumber \\
& = & {1\over 2} (\mu\otimes L_{-1}\mu) + {1\over 2}
(\mu\otimes L_0\mu) \nonumber \\
& \cong & {1\over 2} (\mu\otimes L_{-1}\mu) 
- {1\over 2} \Delta_{1,0}(L_{-1}) (\mu\otimes\mu) 
- {1\over 16} (\mu\otimes \mu) 
\nonumber \\
& = & - {1\over 2} (L_{-1}\mu\otimes\mu) 
- {1\over 16} (\mu\otimes \mu) \,.
\end{eqnarray}
Putting this relation into (\ref{L0calc}) it follows that $L_0$ maps
the space spanned by (\ref{basis}) into itself, and thus that it can
be represented by the matrix 
\be
L_0 = \left( 
\begin{array}{cc} -{1\over 4} & 1 \\
                  -{1\over 16} & {1\over 4}
\end{array}
\right)\,.
\ee
This matrix has
\be
\hbox{tr}(L_0) = 0 \qquad \hbox{det}(L_0) = 0 \,,
\ee
and therefore is conjugate to the matrix
\be\label{jordan}
\left( 
\begin{array}{cc} 0 & 1 \\
                  0 & 0
\end{array}
\right)\,.
\ee
By choosing a suitable basis (for which $L_0$ is of Jordan normal form
as in (\ref{jordan})), we can take the space of ground states to be
spanned by two vectors $\Omega$ and $\omega$ for which the action of
$L_0$ is given as 
\be\label{claimL0}
L_0\, \omega = \Omega \qquad \qquad L_0\, \Omega = 0 \,.
\ee
In fact, $\omega$ and $\Omega$ are given in terms of the two states in 
(\ref{basis}) as 
\begin{eqnarray}
\omega & = & (\mu\otimes\mu) \,, \nonumber \\
\Omega & = & -{1\over 4} (\mu \otimes \mu) 
             + (L_{-1}\mu \otimes \mu) \,. \nonumber
\end{eqnarray}

We now claim that these states can be identified with the states
that appeared in the operator product expansion (\ref{opelog}), thus
demonstrating the equivalence of the two approaches. In order to see
this, let us consider the 3-point function with a suitable $\Omega'$,  
\be
\Bigl\langle \Omega'(\infty)\; \mu(z) \mu(0) \Bigr\rangle 
= z^{1 \over 4} \Bigl(A + B \log (z)\Bigr) \,,
\ee
where $A$ and $B$ are constants (that depend now on $\Omega'$). 
We are interested in the transformation of this amplitude under a
rotation by $2\pi$; this is implemented by the M\"obius
transformation $\exp(2\pi i L_0)$, 
\begin{eqnarray}
\Bigl\langle \Omega'(\infty) \; \e^{2\pi i L_0} 
\Bigl(\mu(z) \mu(0)\Bigr) \Bigr\rangle 
& = & \e^{-{2\pi i \over 4}} 
\Bigl\langle \Omega'(\infty) \; \mu(\e^{2\pi i}z)  \mu(0) \Bigr\rangle 
\nonumber \\
& = & z^{1\over 4} \Bigl(A + B \log(z) + 2\pi i B \Bigr) \,, \label{rech1}
\end{eqnarray}
where we have used that the usual transformation law of 
conformal fields is 
\be
\lambda^{L_0} V(\psi,z) \lambda^{-L_0} = 
V\left( \lambda^{L_0} \psi,\lambda z \right)\,.
\ee
On the other hand, because of (\ref{opelog}) we can rewrite  
\be\label{rech2}
\Bigl\langle \Omega'(\infty) \; \e^{2\pi i L_0} 
\Bigl(\mu(z) \mu(0) \Bigr) \Bigr\rangle 
= z^{1\over 4} \Bigl\langle \Omega'(\infty) \; \e^{2\pi i L_0}
\Bigl(\omega(0) + \log(z) \Omega(0) \Bigr) \Bigr\rangle \,.
\ee
Comparing (\ref{rech1}) with (\ref{rech2}) we then find that 
\begin{eqnarray}
\e^{2 \pi i L_0} \Omega & = & \Omega \\
\e^{2 \pi i L_0} \omega & = & \omega + 2\pi i  \Omega \,,
\end{eqnarray}
\ie\ we reproduce (\ref{claimL0}). 

By analysing various bigger quotient spaces one can actually determine
the structure of the resulting representation (that we shall call
${\cal R}_{1,1}$) in more detail \cite{GKau96a}. One finds that it is
generated from a highest weight state $\omega$ satisfying   
\be
L_0\, \omega = \Omega\,, \qquad L_0\, \Omega = 0\,, \qquad
L_n\, \omega = 0 \quad \hbox{for $n>0$}
\ee
by the action of the Virasoro algebra. The state $L_{-1}\Omega$ is a
null-state of ${\cal R}_{1,1}$, but $L_{-1} \omega$ is not null since  
$L_1 L_{-1} \omega = [L_1,L_{-1}] \omega = 2 L_0 \omega = 2 \Omega$. 
Schematically the representation can therefore be described as 
\begin{center}
  \begin{picture}(180,110)(-10,20)
    \multiput(40,40)(80,0){2}{\vbox to 0pt
      {\vss\hbox to 0pt{\hss$\bullet$\hss}\vss}}
    \put(80,80){\vbox to 0pt
      {\vss\hbox to 0pt{\hss$\bullet$\hss}\vss}}
    \put(0,80){\vbox to 0pt
      {\vss\hbox to 0pt{\hss$\times$\hss}\vss}}
    \put(40,120){\vbox to 0pt
      {\vss\hbox to 0pt{\hss$\times$\hss}\vss}}

    \put(115,40){\vector(-1,0){70}}
    \put(115,45){\vector(-1,1){30}}
    \put(75,75){\vector(-1,-1){30}}

    \multiput(35,45)(-12,12){2}{\line(-1,1){10}}
    \put(11,69){\vector(-1,1){6}}
    \multiput(75,85)(-12,12){2}{\line(-1,1){10}}
    \put(51,109){\vector(-1,1){6}}
    \multiput(35,115)(-12,-12){2}{\line(-1,-1){10}}
    \put(11,91){\vector(-1,-1){6}}

    \put(150,37){$h=0$}       %% or (1, 2t-n)
    \put(150,80){$h=1$}
    \put(36,20){$\Omega$}
    \put(117,20){$\omega$}
    \put(-50,80){${\cal R}_{1,1}$}
  \end{picture}
\end{center}
Here each vertex $\bullet$ denotes a state of the representation
space, and the vertices $\times$ correspond to null-vectors. An arrow
$A\longrightarrow B$ indicates that the vertex $B$ is in the image of
$A$ under the action of the Virasoro algebra. The representation
${\cal R}_{1,1}$ is not irreducible since the states that are obtained
by the action of the Virasoro algebra from $\Omega$ form a
subrepresentation $\H_0 $ of ${\cal R}_{1,1}$ (that is actually
isomorphic to the vacuum representation). On the other hand, 
${\cal R}_{1,1}$ is not completely reducible since we cannot find a
complementary subspace to $\H_0$ that is a representation by itself;
${\cal R}_{1,1}$ is therefore called an {\em indecomposable} (but
reducible) representation. 

The theory at $c=-2$ is not rational (it has infinitely many
representations), but it possesses a preferred class of
`quasirational' representations that are characterised by the property
that they possess a non-trivial null-vector \cite{Nahm}. The simplest 
quasirational representation is the representation generated from
$\mu$ that we have been discussing so far. However, the theory also
has other such quasirational representations; they are generated from
a highest weight state (\ie\ a state that is annihilated by all $L_n$ 
with $n>0$) with $L_0$ eigenvalue 
\be\label{hminimal}
h_{(r,s)} = {(2r - s)^2 - 1 \over 8}  \,,
\ee
where $r$ and $s$ are positive integers. Since the formula for
$h_{r,s}$ has the symmetry  
\be
h_{r,s}=h_{1-r,2-s}=h_{r-1,s-2} \,,
\ee
we can restrict ourselves to the values $(r,s)$ with $s=1,2$. The
representation $\H_{\mu}$ corresponds to the choice $(r,s)=(1,2)$,
while the vacuum representation, ${\cal H}_0$, is $(r,s)=(1,1)$.

By considering fusion products involving any two such quasirational
representations one finds that other indecomposable representation are
generated. These representations can be labelled by $(m,n)$ (where,
for $c=-2$, we always have $n=1$), and their structure is
schematically described as 
\begin{center}
  \begin{tabular}{c@{\hskip0.5in}c}
  \begin{picture}(180,180)(-10,-20)
    \put(80,0){\vbox to 0pt
      {\vss\hbox to 0pt{\hss$\bullet$\hss}\vss}}
    \multiput(40,40)(80,0){2}{\vbox to 0pt
      {\vss\hbox to 0pt{\hss$\bullet$\hss}\vss}}
    \put(80,80){\vbox to 0pt
      {\vss\hbox to 0pt{\hss$\bullet$\hss}\vss}}
    \put(0,80){\vbox to 0pt
      {\vss\hbox to 0pt{\hss$\times$\hss}\vss}}
    \put(40,120){\vbox to 0pt
      {\vss\hbox to 0pt{\hss$\times$\hss}\vss}}

    \put(75,5){\vector(-1,1){30}}
    \put(115,35){\vector(-1,-1){30}}
    \put(115,45){\vector(-1,1){30}}
    \put(75,75){\vector(-1,-1){30}}

    \multiput(35,45)(-12,12){2}{\line(-1,1){10}}
    \put(11,69){\vector(-1,1){6}}
    \multiput(75,85)(-12,12){2}{\line(-1,1){10}}
    \put(51,109){\vector(-1,1){6}}
    \multiput(35,115)(-12,-12){2}{\line(-1,-1){10}}
    \put(11,91){\vector(-1,-1){6}}

    \put(85,-10){$\xi_{m,n}$}
    \put(125,40){$\psi_{m,n}$}
    \put(85,85){$\rho_{m,n}$}
    \put(35,35){\hbox to 0pt{\hss$\phi_{m,n}$}}
    \put(-5,75){\hbox to 0pt{\hss$\phi'_{m,n}$}}
    \put(45,125){$\rho'_{m,n}$}
  \end{picture}
&
  \begin{picture}(180,120)(-10,20)
    \multiput(40,40)(80,0){2}{\vbox to 0pt
      {\vss\hbox to 0pt{\hss$\bullet$\hss}\vss}}
    \put(80,80){\vbox to 0pt
      {\vss\hbox to 0pt{\hss$\bullet$\hss}\vss}}
    \put(0,80){\vbox to 0pt
      {\vss\hbox to 0pt{\hss$\times$\hss}\vss}}
    \put(40,120){\vbox to 0pt
      {\vss\hbox to 0pt{\hss$\times$\hss}\vss}}

    \put(115,40){\vector(-1,0){70}}
    \put(115,45){\vector(-1,1){30}}
    \put(75,75){\vector(-1,-1){30}}

    \multiput(35,45)(-12,12){2}{\line(-1,1){10}}
    \put(11,69){\vector(-1,1){6}}
    \multiput(75,85)(-12,12){2}{\line(-1,1){10}}
    \put(51,109){\vector(-1,1){6}}
    \multiput(35,115)(-12,-12){2}{\line(-1,-1){10}}
    \put(11,91){\vector(-1,-1){6}}

    \put(125,40){$\psi_{1,n}$}
    \put(85,85){$\rho_{1,n}$}
    \put(35,35){\hbox to 0pt{\hss$\phi_{1,n}$}}
    \put(-5,75){\hbox to 0pt{\hss$\phi'_{1,n}$}}
    \put(45,125){$\rho'_{1,n}$}
  \end{picture}
\\
  ${\cal R}_{m,n}$ & ${\cal R}_{1,n}$
  \end{tabular}
\end{center}
The representation ${\cal R}_{m,n}$ is generated from the vector
$\psi_{m,n}$ by the action of the Virasoro algebra, where
\begin{eqnarray} 
L_0 \psi_{m,n} &=& h_{(m,n)} \psi_{m,n} + \phi_{m,n}\,, \\ 
L_0 \phi_{m,n} &=& h_{(m,n)} \phi_{m,n}\,, \\
L_k \psi_{m,n} &=& 0 \qquad \hbox{for $k\geq 2$}\,.
\end{eqnarray}
If $m=1$ we have in addition $L_1\psi_{m,n}=0$, whereas if $m\geq 2$, 
$L_1\psi_{m,n}\ne 0$, and in fact
\be
L_1^{(m-1)(2-n)} \psi_{m,n} = \xi_{m,n} \,,
\ee
where $\xi_{m,n}$ is a Virasoro highest weight vector of conformal
weight $h=h_{(m-1,2-n)}$. The Verma module generated by $\xi_{m,n}$
has a  singular vector of conformal weight 
\begin{eqnarray}
h_{(m-1,2-n)} + (m-1)(2-n) 
& = & {(2(m-1) - (2-n))^2 + 8 (m-1)(2-n) - 1 \over 8} \\
& = & {(2m-n)^2 - 1 \over 8} = h_{(m,n)} \,,
\end{eqnarray}
and this vector is proportional to $\phi_{m,n}$; this singular vector
is however not a null-vector in ${\cal R}_{m,n}$ since it does not
vanish in an amplitude with $\psi_{m,n}$. Again, the representation
${\cal R}_{m,n}$ is reducible since the states generated from $\xi$
form a subrepresentation. On the other hand, it is impossible to find
a complementary subspace that also defines a representation, and
therefore ${\cal R}_{m,n}$ is indecomposable. It was shown in
\cite{GKau96a} that the set of representations that consists of all
quasi-rational irreducible representations and the above
indecomposable representations closes under fusion, \ie\ any fusion
product of two such representations can be decomposed as a direct sum  
of these representations.  

We would like to stress that all but one of these indecomposable
representations (namely the representation ${\cal R}_{1,1}$) are 
{\it not} generated from a highest weight state. This seems to suggest
that generically, logarithmic representations will not be generated
from a highest weight state. We should also mention that it follows
from the structure described above that $\phi$ is necessarily 
null in the subrepresentation generated from $\xi$. In order to see
this, let us denote by $h$ the conformal weight of $\phi$, and let us
assume that $\chi$ is an arbitrary state of conformal weight $h$ in
the subrepresentation generated from $\xi$. (Recall that $L_0$ acts
diagonally on the representation generated from $\xi$.) Then we
find that  
\begin{eqnarray}
\langle \chi \, | \phi \rangle & = & 
\langle \chi \, | (L_0-h) \psi \rangle \nonumber \\
& = & \langle (L_0-h) \chi \, | \psi \rangle = 0 \,,\label{xxx}
\end{eqnarray}
where we have used Dirac notation, \ie\ 
\be
| \phi \rangle = \lim_{z\rightarrow 0} V(\phi,z) \rangle \,, \qquad 
\langle \phi | = \lim_{z\rightarrow\infty} 
z^{2h_{\phi}} \langle V(\phi,z)\,.
\ee
Thus it follows that $\phi$ is orthogonal to any state in the
subrepresentation generated by $\xi$, and therefore that it defines a 
null-state in this (sub)representation. In particular, this implies
that any correlation function that involves $\phi$ as well as only
states from the subrepresentation generated by $\xi$ necessarily
vanishes.  
\newpage

\section{The triplet model}
\renewcommand{\theequation}{3.\arabic{equation}}
\setcounter{equation}{0}

In the previous section we have described in some detail a
`quasirational' theory whose fusion closes on a set of representations
that involves irreducible as well as (logarithmic) indecomposable
representations. For some time it was thought that logarithmic
representations could only occur for non-rational theories; in fact
there was a conjecture in the mathematical literature (due to Dong \&
Mason \cite{DonMas96}) that the $C_2$ condition of Zhu (which implies
in particular that the theory has only finitely many representations)
implies that the theory is `rational' in the strong mathematical
sense, \ie\ that all of the allowed (highest weight) representations
are actually completely decomposable. This has turned out to be wrong,
although the relevant counter example is quite complicated: it
involves the extension of the $c=-2$ model to the {\it triplet
algebra} \cite{Kausch91} that we want to discuss next. (The following
section is largely based on \cite{GKau96b}.)

\subsection{The triplet algebra}

The $c=-2$ Virasoro theory is not rational with respect to the
Virasoro algebra (in particular, the vacuum representation does not
contain any non-trivial null-vectors, and therefore the $C_2$
condition of Zhu is not satisfied), but it is rational with respect to
some extension of the chiral algebra. Let us recall that the conformal
weights of the quasirational representations are given as 
\be
h_{(r,s)} = {(2r - s)^2 - 1 \over 8}  \,,
\ee
and therefore that $h_{3,1}=3$. The triplet algebra is the extension
of the Virasoro theory by a triplet of fields with $h=h_{3,1}=3$ that
we shall denote by $W^i$. The corresponding modes satisfy the
commutation relations 
\begin{eqnarray}
  {}[ L_m, L_n ] &=& (m-n)L_{m+n} - \frac16 m(m^2-1) \delta_{m+n}, 
  \nonumber \\
  {}[ L_m, W^a_n ] &=& (2m-n) W^a_{m+n}, 
  \nonumber \\
  {}[ W^a_m, W^b_n ] &=& g^{ab} \biggl( 
  2(m-n) \Lambda_{m+n} 
  +\frac{1}{20} (m-n)(2m^2+2n^2-mn-8) L_{m+n} 
  \nonumber\\&&\qquad
  -\frac{1}{120} m(m^2-1)(m^2-4)\delta_{m+n}
  \biggr) 
  \nonumber \\&&
  + f^{ab}_c \left( \frac{5}{14} (2m^2+2n^2-3mn-4)W^c_{m+n} 
    + \frac{12}{5} V^c_{m+n} \right)\,, \nonumber
  \end{eqnarray}
where $\Lambda = \mathopen:L^2\mathclose: - 3/10\, \partial^2L$ and 
$V^a = \mathopen:LW^a\mathclose: - 3/14\, \partial^2W^a$ are
quasiprimary normal ordered fields. $g^{ab}$ and $f^{ab}_c$ are the
metric and structure constants of $su(2)$. In an orthonormal basis we
have $g^{ab} = \delta^{ab}, f^{ab}_c = i\epsilon^{abc}$. 

The triplet algebra (at $c=-2$) is only associative, because certain
states in the vacuum representation (which would generically violate
associativity) are null. The relevant null vectors are
\begin{eqnarray} 
  N^a &=& 
  \left(2 L_{-3} W^a_{-3} -\frac43 L_{-2} W^a_{-4} + W^a_{-6}\right)
  \vac\,, 
\label{eq:nulllw}
\\
  N^{ab} &=&
  W^a_{-3} W^b_{-3} \vac - 
  g^{ab} \left( \frac89 L_{-2}^3 + \frac{19}{36} L_{-3}^2 +
  \frac{14}{9} L_{-4}L_{-2} - \frac{16}{9} L_{-6} \right)\vac 
  \nonumber\\*&&
  - f^{ab}_c\left( -2 L_{-2} W^c_{-4} + \frac54 W^c_{-6}
  \right) \vac \,.
\label{eq:nullww}
\end{eqnarray}
Before we begin to discuss the allowed representations of the triplet
algebra, let us first explain, in some more generality, how 
representations of a conformal field theory are constrained by the
structure of the theory. Although the following is presumably well
known by many experts, it does not seem to be widely appreciated.

\subsection{Interlude 1: Representations of a conformal field theory}

Given the vacuum representation of a conformal field theory, we want
to determine which representations are compatible with the vacuum
representation (in a sense that will be explained below). Let us
illustrate the relevant arguments with a simple example, the WZW model 
of $su(2)$ at level $k$ \cite{wzw}. This theory is generated by three
fields of conformal weight $h=1$, $J^\pm$ and $J^3$, whose modes
satisfy the commutation relations 
\begin{eqnarray}
{}[J^3_m,J^3_n] & = & \half k m \delta_{m,-n} \nonumber \\
{}[J^3_m,J^\pm_n] & = & \pm J^{\pm}_{m+n} \label{su2} \nonumber \\
{}[J^+_m,J^-_n] & = &  2 J^3_{m+n} +  k m \delta_{m,-n} \,.
\end{eqnarray}
It defines a conformal field theory since one can construct a stress
energy tensor out of bilinears of the currents $J^a$. The
corresponding modes, $L_m$, define a Virasoro algebra with central
charge 
\be
c = {3 k \over (k+2)} \,,
\ee
and satisfy the commutation relations
\be
[L_m,J^a_n]  =  - n J^a_{m+n} \,.
\ee
The theory is rational provided that $k\in\Nop$. In this case the
irreducible vacuum representation has a null-vector at level $k+1$, 
\be\label{nullstate}
\Nu = \left(J^+_{-1}\right)^{k+1} \Omega \,.
\ee
Indeed, it follows directly from the commutation relations that 
$J^3_n\Nu=J^+_n\Nu=0$ for $n>0$, and for $J^-_n\Nu$ this is a
consequence of $L_n\,\Nu=0$ for $n>0$ together with
\begin{eqnarray}
J^-_1 \Nu & = & \sum_{l=0}^{k} (J^+_{-1})^l [ J^-_1,J^+_{-1} ] 
(J^+_{-1})^{k-l} \Omega \nonumber \\
& = & \left[ k (k+1) - 2 \sum_{l=0}^{k} (k-l) \right](J^+_{-1})^k \Omega 
= 0 \,.
\end{eqnarray}
The zero modes of the affine algebra $\hat{su}(2)$ form the finite Lie
algebra of $su(2)$. For every (finite-dimensional) representation $R$
of $su(2)$, \ie\ for every spin $j\in\Zop/2$, we can construct a Verma 
module for $\hat{su}(2)$ whose Virasoro highest weight space
transforms as $R$. This (and any irreducible quotient space thereof)
defines a representation of the affine algebra $\hat{su}(2)$. However,
not every such representation defines a representation of the
conformal field theory. The additional constraint that every
representation of the conformal field theory has to satisfy is that
it does not modify the structure of the vacuum representation (see
\cite{GabGod} for a more formal discussion of this point.) More
precisely, if $\phi_i$ are states in representations of the conformal
field theory, then any correlation function involving these fields
together with a null state of the vacuum representation must vanish, 
\be\label{crucial}
\langle \phi_1(z_1) \cdots \phi_n(z_n) V(\Nu,u) \rangle = 0 \,,
\ee
where $\Nu$ is for example the null-state (\ref{nullstate}). [If this
was not the case, then $\Nu$ would not be a null state in the whole
theory, and therefore the structure of the vacuum representation would
have been modified by the `representations' $\phi_i$.] One particular
case of (\ref{crucial}) arises when $n=2$, and $\phi_1$ and $\phi_2$
are in conjugate representations. Then we can write (\ref{crucial}) as 
\be
\langle \phi_1 | V(\Nu,u) | \phi_2 \rangle = 0  \,.
\ee
We can expand this condition in terms of modes; the first non-trivial
condition is then simply that the zero mode of the null-vector 
$\Nu$ annihilates every allowed state in the representation (see also
\cite{FNO92}). It is most convenient to evaluate this condition on a
highest weight state; in this case the zero mode of $\Nu$ can be
easily calculated, and one finds that  
\be
\label{nulleq}
V_0(\Nu) \phi = \left(J^+_0\right)^{k+1} \phi \,.
\ee
Thus in order for the $\hat{su}(2)$ representation generated from
$\phi$ to define a representation of the conformal field theory,
(\ref{nulleq}) must vanish. This implies that $j$ can only take the
values $j=0,1/2,\ldots,k/2$. Since $\Nu$ generates all other
null-fields, one may suspect that this is the only additional
condition, and this is indeed correct \cite{FreZhu92}.   

Incidentally, in the case at hand the vacuum theory is actually
unitary, and the allowed representations are precisely those
representations of the affine algebra that are unitary with respect to
an inner product for which
\be
\left( J^\pm_n \right)^\dagger = J^\mp_{-n} \qquad
\left( J^3_m \right)^\dagger = J^3_{-m} \,.
\ee
Indeed, if $|j,j\rangle$ is a Virasoro highest weight state with
$J^+_0|j,j\rangle=0$ and  $J^3_0 |j,j\rangle=j |j,j\rangle$, then  
\begin{eqnarray}
\Bigl( J^+_{-1} |j,j\rangle, J^+_{-1}|j,j\rangle \Bigr) & =  &
\Bigl( |j,j\rangle, J^-_1 J^+_{-1} |j,j\rangle \Bigr) \nonumber \\
& = & (k-2j) \Bigl( |j,j\rangle, |j,j\rangle \Bigr) \,,
\end{eqnarray}
and if the representation is unitary, this requires that $(k-2j)\geq 0$, 
and thus that $j=0,1/2,\ldots, k/2$. As it turns out, this is also
sufficient to guarantee unitarity. In general, however, the
constraints that select the representations of the conformal field
theory from those of the Lie algebra of modes cannot be understood in
terms of unitarity. (In particular, this is not possible for the
non-unitary minimal models where unitarity does not seem to play any
role.)

\subsection{The representations of the triplet algebra}

Let us now return to the question of determining the allowed
representations of the triplet theory. We are only interested in
representations for which the spectrum of $L_0$ is bounded from
below, and which therefore possess a highest weight state (although we
do not assume that the whole representation is generated from this
state by the action of the modes). As we have explained above, the
zero modes of the null-states have to vanish on the highest weight
states, and this will restrict the allowed representations. Evaluating
the constraint coming from (\ref{eq:nullww}), we find 
\begin{equation}\label{3}
  \left(W^a_0 W^b_0 - g^{ab} \frac19 L_0^2 (8L_0 + 1) - 
  f^{ab}_c \frac15 (6L_0-1) W^c_0 \right) \psi = 0 \,,
\label{eq:wwzero}
\end{equation}
where $\psi$ is any highest weight state, while the relation coming
from the zero mode of (\ref{eq:nulllw}) is satisfied identically. 
Furthermore, the constraint from $W^a_1 N^{bc}_{-1}$, together with
(\ref{eq:wwzero}) implies that $W^a_0 (8 L_0 - 3) (L_0 -1) \psi = 0$. 
Multiplying with $W_0^a$ and using (\ref{eq:wwzero}) again, this
implies that 
\begin{equation}
\label{eq:heigen}
0  = L_0^2 (8 L_0 + 1) (8 L_0 - 3) (L_0 - 1) \psi\,.
\end{equation}
For irreducible representations, $L_0$ has to take a fixed value $h$
on the highest weight states, and (\ref{eq:heigen}) then implies that
$h$ has to be either $h=0, -1/8, 3/8$ or $h=1$. However, it also
follows from (\ref{eq:heigen}) that a logarithmic highest weight 
representation is allowed since we only have to have that $L_0^2=0$
but not necessarily that $L_0=0$. Thus in particular, a
two-dimensional space of highest weight states with relations
\be\label{3.16}
L_0\, \omega = \Omega \qquad \qquad L_0\, \Omega = 0 \,.
\ee
satisfies (\ref{eq:heigen}). 

As we shall see this is {\it not} the only logarithmic representation
that will play a role for this theory, but it is the only logarithmic
representation that is generated from a highest weight state, \ie\
from a state $\phi$ for which $V_n(\psi)\phi=0$ for all $n>0$. This
property was assumed in the derivation of (\ref{eq:heigen}), and it is
therefore not surprising that the other logarithmic representation has
not been detected by this analysis.

Returning to the classification of highest weight representations, it
also follows from (\ref{eq:wwzero}) that 
\begin{displaymath}
  {}[ W^a_0, W^b_0 ] = \frac25 (6h-1) f^{ab}_c W^c_0 \,,
\end{displaymath}
which is a rescaled version of the $su(2)$ algebra. After rescaling, 
the irreducible representations of these zero modes can then be
labelled by $j$ and $m$, where $j(j+1)$ is the  eigenvalue of the
Casimir operator $\sum_a (W^a_0)^2$, and $m$ is the eigenvalue of
$W^3_0$. Because of (\ref{eq:wwzero}), $W^a_0 W^a_0 = W^b_0
W^b_0$ on the highest weight states, and thus $j(j+1)=3m^2$. This can 
only be satisfied for $j=0, 1/2$, and this restricts the allowed
representations to \cite{EHHu93,GKau96b}
\begin{itemize}
\item the singlet representations, $j=0$, at $h=0, -1/8$,
\item the doublet representations, $j=1/2$, at $h=1, 3/8$.
\end{itemize}

The above analysis is really a physicist's way of analysing
(irreducible) representations of {\it Zhu's algebra} \cite{Zhu}. Since
Zhu's construction is a powerful and important technique, we want to
use the opportunity to give a brief explanation of his work.

\subsection{Interlude 2: Zhu's algebra}

The above analysis suggests that to each representation of the zero
modes of the fields for which the zero modes of the null-fields
vanish, a highest weight representation of the conformal field  
theory can be associated, and that all highest weight representations
of a conformal field theory can be obtained in this way
\cite{EFHHNV92}. This idea has been made precise by Zhu \cite{Zhu} who
constructed an algebra, now commonly referred to as 
{\it Zhu's algebra}, that describes the algebra of zero modes modulo
zero modes of null-vectors, and whose representations are in
one-to-one correspondence with those of the conformal field
theory. The following explanation of Zhu's work follows closely   
\cite{GabGod}. 

In a first step we determine the subspace of states whose zero modes
always vanish on Virasoro highest weight states. This subspace
certainly contains the states of the form $(L_{-1}+L_0)\psi$, where
$\psi$ is an arbitrary state in the vacuum representation $\H_0$.
Indeed, it follows from (\ref{deri1}) together with the mode expansion
of an arbitrary chiral field, 
\be
V(\psi,z) = \sum_{n\in\Zop} V_n(\psi) z^{-n-h_\psi} 
\ee
that 
\be
\label{derimode}
V_{n}(L_{-1}\psi) = - (n+h) V_{n}(\psi) \,,
\ee
and thus
\be
\label{null1}
V_0 \left( (L_{-1} + L_0) \psi \right) = V_0\left(L_{-1}\psi\right)
+ h V_0\left(\psi\right) = 0 \,.
\ee
Furthermore, the subspace must also contain every state whose zero
mode is of the form $V_0(\chi) V_0((L_{-1}+L_0)\psi)$ or
$V_0((L_{-1}+L_0)\psi) V_0(\chi)$. In order to describe states of this
form more explicitly, it is useful to observe that if both $\phi$ and
$\bar\phi$ are Virasoro highest weight states 
\begin{eqnarray}
\langle \bar\phi | V(\psi,1) |\phi\rangle & = &
\sum_l \langle \bar\phi | V_l(\psi) |\phi\rangle \nonumber \\
& = & \langle \bar\phi| V_0(\psi) | \phi\rangle \,, \label{zhu1}
\end{eqnarray}
since the highest weight property implies that
$V_l(\psi)|\phi\rangle=0$ for $l>0$ and similarly
$\langle\bar\phi| V_l(\psi)=0$ for $l<0$. Because of the translation
symmetry of the amplitudes, this can then be rewritten as  
\be
\label{zhu2}
\langle \bar\phi | \phi(-1)\; \psi\rangle = 
\langle \bar\phi| V_0(\psi)| \phi\rangle \,.
\ee
Let us introduce the operators
\be
\label{zhu3}
V^{(N)}(\psi) = \oint_{0} {d w \over w^{N+1}} 
V\left( \left(w+1 \right)^{L_0} \psi, w \right) \,,
\ee
where $N$ is an arbitrary integer, and the contour is a small circle
that encircles $w=0$ but not $w=-1$. Then if both $\phi$ and
$\bar\phi$ are arbitrary Virasoro highest weight states and $N>0$, we
have that 
\be
\label{zhu4}
\langle \bar\phi | \phi(-1) \;V^{(N)}(\psi)\chi\rangle = 0 \,,
\ee
since the integrand in (\ref{zhu4}) does not have any poles at $w=-1$
or $w=\infty$. Here we have used that, because of the highest weight
property of $\phi$, the leading singularity in the operator product
expansion is 
\be
V(\psi,w) \, V(\phi,z) \sim {1\over (w-z)^{h_\psi}} 
\Bigl(V(V_0(\psi)\phi,z) + O(w-z) \Bigr)\,,
\ee
and similarly for $\bar\phi$. Because of (\ref{zhu2}) it now follows 
that the zero mode of the state $V^{(N)}(\psi)\chi$ with $N>0$
vanishes on an arbitrary highest weight state (since the amplitude
with any other highest weight state vanishes). Let us denote by
$O(\H_0)$ the subspace of $\H_0$ that is generated by states of the
form $V^{(N)}(\psi)\chi$ with $N>0$, and define the quotient space
$\A(\H_0)=\H_0 / O(\H_0)$. The above then implies that we can
associate a zero mode (acting on a highest weight state) to each state
in $\A(\H_0)$. We can write (\ref{zhu3}) in terms of modes as  
\be
\label{zhumode}
V^{(N)}(\psi) = \sum_{n=0}^{h} {h\choose n} V_{-n-N}(\psi) \,,
\ee
where $\psi$ has conformal weight $h$, and it therefore follows that
\be
V^{(1)}(\psi)\Omega = V_{-h-1}(\psi)\Omega + h V_{-h}(\psi)\Omega 
                  = (L_{-1}+L_0) \psi \,.
\ee
Thus $O(\H_0)$ contains the states in (\ref{null1}). Furthermore, 
\begin{eqnarray}
V^{(N)}(L_{-1}\psi)  & = &
\oint_{0} {d w \over w^{N+1}} \left( w +1 \right)^{h_\psi+1}
{dV( \psi, w ) \over dw} \nonumber \\
& = & -\oint_{0} dw{d\over dw}\left(
{( w +1)^{h_\psi+1} \over w^{N+1}}\right)
V( \psi, w ) \nonumber \\
& = &  (N-h_\psi)V^{(N)}(\psi) + (N+1)V^{(N+1)}(\psi)  \,,
\end{eqnarray}
and this implies, for $N\ne -1$, 
\be
\label{recursive}
V^{(N+1)}(\psi) = {1\over N+1} V^{(N)}(L_{-1}\psi) 
- {N-h_\psi\over N+1}V^{(N)}(\psi)\,.
\ee 
Thus $O(\H_0)$ is actually generated by the states of the form
$V^{(1)}(\psi)\chi$, where $\psi$ and $\chi$ are arbitrary states in
$\H_0$. 

As we shall show in the appendix, the vector space $\A(\H_0)$ actually
has the structure of an associative algebra, where the product is
defined by  
\be
\label{prodd}
\psi \ast_L \chi \equiv V^{(0)}(\psi) \chi \,,
\ee
and $V^{(0)}(\psi)$ is given as in (\ref{zhu3}) or (\ref{zhumode});
this algebra is called Zhu's algebra. The analogue of (\ref{zhu4}) is
then 
\begin{eqnarray}
\label{zhuaction}
\langle \bar\phi | \phi(-1) \; V^{(0)}(\psi)\chi\rangle & = &
(-1)^{h_\psi} 
\langle V_0(\psi) \bar\phi | \phi(-1) \; \chi\rangle \nonumber \\
& = & (-1)^{h_\psi} 
\langle V_0(\psi) \bar\phi |V_0(\chi) | \phi \rangle \\
& = & \langle \bar\phi | V_0(\psi) V_0(\chi) | \phi \rangle\,,
\end{eqnarray}
and thus the product in $\A(\H_0)$ corresponds indeed to the action of
the zero modes. 

As we mentioned before, this algebra plays the role of the algebra of
zero modes. Since it has been constructed in terms of the space of
states of the vacuum representation, all null-relations have been taken
into account, and one may therefore expect that its irreducible
representations are in one-to-one correspondence with the irreducible
representation of the conformal field theory. This is
indeed true \cite{Zhu}, although the proof is rather non-trivial. 

We should mention that there is a natural generalisation of this
construction for $n$-point functions \cite{GabGod}: one can construct 
a suitable quotient space of the vacuum representation that describes
the different sets of $n$-point functions involving $n$ highest weight
states. While this is probably not an efficient tool for the actual
calculation of $n$-point functions, it is useful from a conceptual
point of view. For example, using this idea, one can show that the
$C_2$ condition of Zhu implies that the theory only has finitely many 
$n$-point functions \cite{GN}. 

There is also a similar quotient space $A_q(\H_0)$ that characterises
the torus amplitudes of the theory. Using this description, it was
shown by Zhu that for a rational theory (in the strong mathematical
sense) for which the $C_2$ condition is satisfied, the characters of
the irreducible representations transform into one another under the
modular transformation, \ie\ 
\be
\chi_i\left(-{1\over \tau}\right) = \sum_j S_{ij}\,
\chi_j(\tau) \,,
\ee
where $S_{ij}$ is a {\it constant} matrix. As was explained by Flohr
in his lectures \cite{FL} (see also below), for the case of the
triplet theory $S_{ij}$ are not constants, but depend on $\tau$. Since
the $C_2$ condition is satisfied, it thus follows that the triplet
theory is not a rational theory in the strong mathematical sense. In
fact, as we have seen above (and will see more below) it does contain
logarithmic (indecomposable) representations!  

It is easy to see (and in fact shown by Zhu) that Zhu's algebra must
be semisimple if the theory is rational in the strong mathematical
sense. This suggests that logarithmic theories may be characterised by
the property that Zhu's algebra is not semisimple.\footnote{This
conjecture arose in discussions with Peter Goddard.} This is certainly
the case for the example of the triplet algebra; it would be
interesting to show that this is true in general.

\subsection{Interlude 2 (ctd): Zhu's algebra for the Yang-Lee model} 

In order to give some idea of the structure of Zhu's algebra, let us
now work out one simple case in detail. The model we want to consider
is the Virasoro minimal model with $c=-{22\over 5}$. In terms of the
discrete series of minimal models where the central charge is
parametrised by  
\be
c=c_{p,q} = 1 - {6 (p-q)^2 \over pq} \,,
\ee
this model corresponds to $(p,q)=(2,5)$. For every such model, the
vacuum representation $\H_0$ has a non-trivial null-vector at level
$(p-1)(q-1)$; for the $(2,5)$ model, we therefore have a null-vector
at level four, and it is explicitly given as 
\be\label{YLnull}
\Nu = \left( L_{-4} - {5\over 3} L_{-2}^2 \right) \Omega \,.
\ee
For the case of the Virasoro field, the Zhu modes are 
\begin{eqnarray}
V^{(0)}(L) & = & (L_0 + 2 L_{-1} + L_{-2}) \nonumber \\
V^{(1)}(L) & = & (L_{-1} + 2 L_{-2} + L_{-3}) \nonumber \\
V^{(2)}(L) & = & (L_{-2} + 2 L_{-3} + L_{-4})\,,
\end{eqnarray}
as follows from (\ref{zhumode}). In order to determine the quotient
space $A(\H_0)$ we can discard any state of the form $V^{(N)}(L)\psi$
with $N>0$. In particular, we can therefore replace any state of the
form $L_{-n}\psi$ with $n\geq 3$, in terms of a linear combination of
states $L_{-m}\psi$ with $m<n$. Repeating this algorithm (and using
the fact that a spanning set for $\H_0$ can be taken to consist of the
states 
\be
L_{-n_1} \cdots L_{-n_m} \Omega \,,
\ee
where $n_1\geq n_2 \geq \cdots \geq n_m \geq 2$), we can thus take
$A(\H_0)$ to be spanned by the states 
\be\label{spanning}
L_{-2}^l \Omega \,,
\ee
where $l=0,1,2,\ldots$. Finally, because of the null vector at level
four (\ref{YLnull}), we can restrict (\ref{spanning}) to $l=0,1$. Thus
Zhu's algebra is two-dimensional in our case.

In order to work out the algebra relation, we only have to calculate
\begin{eqnarray}
L \ast L & = & V^{(0)}(L) L_{-2} \Omega  \nonumber \\
& = & \left(L_{-2}^2 + 2 L_{-1} L_{-2} + L_0 L_{-2}\right) 
\Omega \nonumber \\
& = & \left({3\over 5} L_{-4} + 2 L_{-3} + 2 L_{-2} \right) \Omega \,,
\end{eqnarray}
where we have used the null vector relation (\ref{YLnull}) as well as
$L_{-1}\Omega=0$ in the last line. Next we use the fact that we can
discard states of the form $V^{(N)}(L)\psi$ with $N>0$ to write 
\begin{eqnarray}
L \ast L & \cong & 
\left({3\over 5} L_{-4} + 2 L_{-3} + 2 L_{-2} \right) \Omega  
- {3\over 5} V^{(2)}(L)\Omega \nonumber \\
& = & \left({3\over 5} L_{-4} + 2 L_{-3} + 2 L_{-2} \right) \Omega   
- {3\over 5} \left( L_{-4} + 2 L_{-3} + L_{-2} \right) \Omega
\nonumber \\
& = & \left({4\over 5} L_{-3} + {7\over 5} L_{-2} \right) \Omega 
\nonumber \\
& \cong &  \left({4\over 5} L_{-3} + {7\over 5} L_{-2} \right) \Omega 
- {4\over 5} V^{(1)}(L)\Omega 
\nonumber \\
& = & \left({4\over 5} L_{-3} + {7\over 5} L_{-2} \right) \Omega 
- {4\over 5} \left( L_{-3} + 2 L_{-2} + L_{-1} \right) \Omega
\nonumber \\
& = & - {1\over 5} L_{-2} \Omega \,.
\end{eqnarray}
Thus we have shown that in Zhu's algebra we have the relation
\be
L \ast L + {1\over 5} L = L \ast \left(L+{1\over 5}\right) = 0 \,.
\ee
The only allowed highest weight representations are therefore 
\begin{eqnarray}
h & = & 0  \qquad\qquad \hbox{(vacuum)} \nonumber \\
h & = & -{1\over 5}\,.  
\end{eqnarray}
This reproduces precisely the representations in the Kac table for the
$(2,5)$ model: as was explained by Flohr in his lectures \cite{FL},
the relevant conformal weights are 
\be
h_{r,s} = {(rq -sp)^2 - (p-q)^2 \over 4 pq}
\ee
where $1\leq r \leq p-1$, $1\leq s \leq q-1$ and 
$h_{r,s} = h_{p-r,q-s}$.  For $(p,q)=(2,5)$, we therefore have 
$h_{1,1}=h_{1,4}=0$ and $h_{1,2}=h_{1,3}=-{1\over 5}$. 

More generally, as we have mentioned above the vacuum representation
for $c=c_{p,q}$ has a null-vector at level $N=(p-1)(q-1)$. 
Furthermore, we can always take $A(\H_0)$ to be spanned by
$L_{-2}^l\Omega$ where $l=0,1,\ldots$. The coefficient of the term 
$L_{-2}^{N/2}\Omega$ in the null-vector at level $N=(p-1)(q-1)$ is
non-trivial, and therefore, $A(\H_0)$ is spanned by $L_{-2}^l\Omega$
with $l=0,1,\ldots,(p-1)(q-1)/2-1$. Thus we have
\be
\dim(A(\H_0)) = {(p-1)(q-1) \over 2} \,,
\ee
and this is in agreement with the number of representations in the Kac
table. In fact, the algebra structure of $A(\H_0)$ is given as
\cite{Wang} 
\be
A(\H_0) = P[L] / \prod_{(r,s)} (L-h_{r,s})\,,
\ee
where $P[L]$ is the polynomial algebra in the variable $L$, and the
product in the denominator runs over the set of inequivalent
representations in the Kac table, \ie\ $1\leq r \leq p-1$, 
$1\leq s \leq q-1$  and we have identified $(r,s)$ with $(p-r,q-s)$.

\subsection{The fusion rules of the triplet theory}

Let us now return to the example of the triplet theory. As we have
seen above, the analysis of Zhu's algebra implies that the theory has
four irreducible representations:
\begin{itemize}
\item the singlet representation at $h=0$, ${\cal V}_0$;
\item the singlet representation at $h=-1/8$, ${\cal V}_{-1/8}$;
\item the doublet representation at $h=1$, ${\cal V}_1$;
\item the doublet representation at $h=3/8$, ${\cal V}_{3/8}$.
\end{itemize}
We can work out the fusion products of these representations using the
algorithm we described in section~2. We find that the fusion products
involving ${\cal V}_0$ are trivial, and for ${\cal V}_1$ we obtain
\cite{GKau96b} 
\begin{equation}
  ({\cal V}_{-1/8}\otimes{\cal V}_{1})_{\mathrm{f}} 
          = {\cal V}_{3/8} \,, \qquad 
  ({\cal V}_{3/8}\otimes{\cal V}_{1})_{\mathrm{f}} 
          = {\cal V}_{-1/8}\,.
\end{equation}
Furthermore we find 
\begin{equation}
\begin{array}{rcl}
  \left({\cal V}_{1}\otimes{\cal V}_{1}\right)_{\mathrm{f}} &=& 
    {\cal V}_0\,, \\
  \left({\cal V}_{-1/8}\otimes{\cal V}_{-1/8}\right)_{\mathrm{f}} &=& 
    {\cal R}_{0}\,, \\
  \left({\cal V}_{-1/8}\otimes{\cal V}_{3/8}\right)_{\mathrm{f}} &=& 
    {\cal R}_{1}\,, \\
  \left({\cal V}_{3/8}\otimes{\cal V}_{3/8}\right)_{\mathrm{f}} &=& 
    {\cal R}_{0}\,,
\end{array}
\end{equation}
where ${\cal R}_{0}, {\cal R}_{1}$ are generalised highest weight
representations whose structure can (schematically) be described by
the following diagram:
\begin{displaymath}
  \begin{array}{c@{\qquad\qquad}c}
    \begin{picture}(150,120)(-10,-20)
      \put(0,0){\vbox to 0pt
        {\vss\hbox to 0pt{\hss$\bullet$\hss}\vss}}
      \put(129,0){\vbox to 0pt
        {\vss\hbox to 0pt{\hss$\bullet$\hss}\vss}}
      \put(40,60){\vbox to 0pt
        {\vss\hbox to 0pt{\hss$\bullet$\hss}\vss}}
      \put(89,60){\vbox to 0pt
        {\vss\hbox to 0pt{\hss$\bullet$\hss}\vss}}
      \put(37,56){\vector(-2,-3){34}}
      \put(85,57){\vector(-3,-2){81}}
      \put(125,3){\vector(-3,2){81}}
      \put(126,5){\vector(-2,3){34}}
      \put(124,0){\vector(-1,0){119}}
      \put(-5,-15){$\Omega$}
      \put(124,-15){$\omega$}
      \put(35,70){$\bpsi_1$}
      \put(84,70){$\bpsi_2$}
    \end{picture}
    &
    \begin{picture}(140,120)(-10,-20)
      \put(0,60){\vbox to 0pt
        {\vss\hbox to 0pt{\hss$\bullet$\hss}\vss}}
      \put(124,60){\vbox to 0pt
        {\vss\hbox to 0pt{\hss$\bullet$\hss}\vss}}
      \put(60,0){\vbox to 0pt
        {\vss\hbox to 0pt{\hss$\bullet$\hss}\vss}}
      \put(64,0){\vbox to 0pt
        {\vss\hbox to 0pt{\hss$\bullet$\hss}\vss}}
      \put(119,60){\vector(-1,0){114}}
      \put(57,3){\vector(-1,1){54}}
      \put(121,57){\vector(-1,-1){54}}
      \put(-5,70){$\bpsi$}
      \put(119,70){$\bphi$}
      \put(50,-15){$\xi^+,\xi^-$}
    \end{picture}
    \\
    {\cal R}_0 & {\cal R}_1
  \end{array}
\end{displaymath}
Here each vertex represents an irreducible representation, 
${\cal V}_0$ in the bottom row and ${\cal V}_1$ in the top row. 
An arrow $A\longrightarrow B$ indicates that the vertex $B$ is in the
image of $A$ under the action of the triplet algebra. Let us describe
the two representations in some more detail.

\noindent {\bf The representation ${\cal R}_0$} is generated from a
highest weight vector $\omega$ of $h=0$ forming a Jordan cell with 
$\Omega$. It is an  extension of the vacuum representation, and its
defining relations are   
\begin{eqnarray}
  L_0\, \omega &=& \Omega\,, \qquad  L_0\, \Omega = 0 \nonumber \\
  L_n\, \omega &=& 0 \,, \qquad \hbox{for $n>0$} \nonumber \\  
  W^a_0 \omega &=& 0\,, \nonumber \\
  W^a_n \omega &=& 0 \,, \qquad \hbox{for $n>0$.} \nonumber
\end{eqnarray}
As we have seen before this representation is compatible with the
vacuum representation (see (\ref{3.16})). The two ground states are
singlets under the action of $W^a_0$, but at higher levels there are
also Jordan cells for $W^a_0$.

\noindent {\bf The representation ${\cal R}_1$} is generated from a
doublet $\phi^\pm$ of cyclic states of weight $h=1$. It has two ground
states  $\xi^\pm$ at $h=0$ and another doublet $\psi^\pm$ at $h=1$
forming $L_0$ Jordan cells with $\phi^\pm$. The defining relations are 
\begin{displaymath}
  \begin{array}{rcl@{\qquad}rcl}
    L_0 \bpsi &=& \bpsi\,, &
    W^a_0 \bpsi &=& 2t^a \bpsi\,,  \\[\bigskipamount]
    L_0 \bxi &=& 0\,, &
    W^a_0 \bxi &=& 0\,, \\
    L_{-1} \bxi &=& \bpsi\,, &
    W^a_{-1} \bxi &=& t^a \bpsi\,, \\[\bigskipamount]
    L_1 \bphi &=& -\bxi\,, &
    W^a_1 \bphi &=& -t^a \bxi\,, \\
    L_0 \bphi &=& \bphi + \bpsi\,, &
    W^a_0 \bphi &=& 2t^a \bphi\,.
  \end{array}
\end{displaymath}
We stress that $\psi^\pm$ and $\phi^\pm$ form a Jordan cell with
respect to $L_0$ but that they remain uncoupled with respect to
$W^a_0$. On higher levels there are also Jordan cells for $W^a_0$. 

It is clear from these relations (as well as the above diagram) that
${\cal R}_1$ is not a highest weight representation: the states
$\phi^\pm$ from which it is generated by the action of the modes are
not highest weight states. Since $\phi^\pm$ are not highest weight,
they do not have to satisfy the relation (\ref{eq:heigen}), and thus
there is no contradiction with the fact that ${\cal R}_1$ is
compatible with the vacuum representation. (An analysis similar to 
what was described in section~3.3 can be performed, and one can show
that $V_0(N^a)\phi^\pm = 0$ and $V_0(N^{ab})\phi^\pm = 0$;
however, since $\phi^\pm$ is not highest weight, the analysis is
slightly more complicated than what was described in section~3.3.)

Finally, we can calculate the fusion products involving the
indecomposable representations. 
\begin{eqnarray}
  \left({\cal V}_{-1/8}\otimes{\cal R}_{0}\right)_{\mathrm{f}} &=& 
    2{\cal V}_{-1/8} \oplus 2{\cal V}_{3/8}\,, \nonumber \\
  \left({\cal V}_{-1/8}\otimes{\cal R}_{1}\right)_{\mathrm{f}} &=& 
    2{\cal V}_{-1/8} \oplus 2{\cal V}_{3/8}\,, \nonumber \\
  \left({\cal V}_{3/8}\otimes{\cal R}_{0}\right)_{\mathrm{f}} &=& 
    2{\cal V}_{-1/8} \oplus 2{\cal V}_{3/8}\,, \nonumber \\
  \left({\cal V}_{3/8}\otimes{\cal R}_{1}\right)_{\mathrm{f}} &=& 
    2{\cal V}_{-1/8} \oplus 2{\cal V}_{3/8}\,, \nonumber
\end{eqnarray}
\begin{eqnarray}
  \left({\cal V}_{1}\otimes{\cal R}_{0}\right)_{\mathrm{f}} &=& 
    {\cal R}_{1}\,, \nonumber \\
  \left({\cal V}_{1}\otimes{\cal R}_{1}\right)_{\mathrm{f}} &=& 
    {\cal R}_{0}\,, \nonumber \\[\bigskipamount]
  \left({\cal R}_{0}\otimes{\cal R}_{0}\right)_{\mathrm{f}} &=& 
    2{\cal R}_{0} \oplus 2{\cal R}_{1}\,, \nonumber \\
  \left({\cal R}_{0}\otimes{\cal R}_{1}\right)_{\mathrm{f}} &=& 
    2{\cal R}_{0} \oplus 2{\cal R}_{1}\,, \nonumber \\
  \left({\cal R}_{1}\otimes{\cal R}_{1}\right)_{\mathrm{f}} &=& 
    2{\cal R}_{0} \oplus 2{\cal R}_{1}\,. \nonumber
\end{eqnarray}
It follows from these results that the set of representations 
${\cal V}_{0}, {\cal V}_{1}, {\cal V}_{-1/8}, {\cal V}_{3/8}$ and 
${\cal R}_{0}, {\cal R}_{1}$, is closed under fusion. The triplet 
algebra at $c=-2$ defines therefore a rational logarithmic conformal
field theory.

\subsection{Characters and modular transformations}

The characters of the irreducible representations of the triplet
algebra have been calculated in \cite{Kau95} (see also
\cite{Flohr95}). From these, and the explicit description of the
indecomposable representations, we can derive the characters of all
the above representations. In more detail we have
\begin{eqnarray}\label{characters}
  \chi_{{\cal V}_0}(\tau) &=&
          \frac12\left(\eta(\tau)^{-1}\theta_{1,2}(\tau) +
              \eta(\tau)^2\right)\,, \nonumber \\ 
  \chi_{{\cal V}_1}(\tau) &=& 
          \frac12\left(\eta(\tau)^{-1}\theta_{1,2}(\tau) -
              \eta(\tau)^2\right)\,, \nonumber \\ 
  \chi_{{\cal V}_{-1/8}}(\tau) &=& \eta(\tau)^{-1}
\theta_{0,2}(\tau)\,, \nonumber  \\
  \chi_{{\cal V}_{3/8}}(\tau) &=& \eta(\tau)^{-1}
\theta_{2,2}(\tau)\,, \\
  \chi_{{\cal R}_0}(\tau) &=& 2\eta(\tau)^{-1} \theta_{1,2}(\tau)\,,
\nonumber \\
  \chi_{{\cal R}_1}(\tau) &=& 2\eta(\tau)^{-1}
\theta_{1,2}(\tau)\,. \nonumber  
\end{eqnarray}
Here the $\theta_{m,k}$-functions are defined as 
\be
\theta_{m,k}(\tau) = \sum_{j\in\Zop + {m\over 2k}} q^{k\, j^2} 
\,,\qquad q=\e^{2\pi i \tau} \,,
\ee
and $\eta(\tau)$ is the Dedekind eta-function,
\be
\eta(\tau) = q^{{1\over 24}} \prod_{n=1}^{\infty} (1-q^n) \,.
\ee
It follows directly from the definition of the $\theta_{m,k}$
functions that we have the identity
\be
\theta_{1,2}(\tau) = \theta_{3,2}(\tau)\,.
\ee
Under the $S$ transformation, $S:\tau\mapsto -1/\tau$, the
$\theta_{m,k}$ functions transform as \cite{kac}
\be\label{thetatrans}
\theta_{m,k}\left(-{1\over \tau}\right) = 
\sqrt{{-i\tau \over 2 k }} \sum_{l\in\Zop\, mod\, 2k\Zop}
\e^{-\pi i l {m\over k}} \theta_{l,k}(\tau) \,,
\ee
while the Dedekind eta-function becomes
\be\label{etatrans}
\eta\left(-{1\over \tau}\right) = \sqrt{-i\tau}\; \eta(\tau)\,.
\ee
It therefore follows that the space generated by the last four
characters in (\ref{characters}) is invariant under the action of the
modular group, and that each has a suitable transformation property
under $S$. On the other hand the $S$ transformation of 
$\chi_{{\cal V}_0}(\tau)$ and  $\chi_{{\cal V}_1}(\tau)$ involves
coefficients which are themselves functions of $\tau$, {\it i.e.}  
\begin{eqnarray*}
  \chi_{{\cal V}_0}(-1/\tau) &=& \frac14 \chi_{{\cal V}_{-1/8}}(\tau) - 
  \frac14 \chi_{{\cal V}_{3/8}}(\tau) - \frac{i\tau}{2}
         \eta(\tau)^2\,, 
  \\
  \chi_{{\cal V}_1}(-1/\tau) &=& \frac14 \chi_{{\cal V}_{-1/8}}(\tau) - 
  \frac14 \chi_{{\cal V}_{3/8}}(\tau) + \frac{i\tau}{2} \eta(\tau)^2\,.  
\end{eqnarray*}
As Flohr has explained in his lectures \cite{FL} (see also
\cite{Flohr95}) it is  tempting to interpret this in terms of a
conventional modular matrix relating generalised characters. Given
Zhu's work it seems plausible that these generalised characters have
an interpretation in terms of torus amplitudes, and it would be very
interesting to understand this more precisely. 

The representations ${\cal R}_0$ and ${\cal R}_1$ contain the
irreducible subrepresentations ${\cal V}_0$ and ${\cal V}_1$,
respectively, and the set of representations 
${\cal V}_{-1/8}, {\cal   V}_{3/8}$ and ${\cal R}_{0}, {\cal R}_{1}$
is already closed under fusion. This suggests that the fundamental
building blocks of the theory are these four representations 
({\it c.f.} also \cite{Roh}). Two of its characters are the same,
and so the definition of the modular matrices is ambiguous. It turns
out that there is a one parameter freedom to define these matrices so
that the relations of the modular group, $S^4=\bbbone$ and
$(ST)^3=S^2$ are satisfied,\footnote{All different choices lead to
equivalent representations of the modular group.} and a unique
solution for which the charge conjugation matrix $S^2$ is a
permutation matrix. In the basis of 
$\chi_{{\cal R}_0}, \chi_{{\cal R}_1}, 
\chi_{{\cal V}_{-1/8}}, \chi_{{\cal V}_{3/8}}$ this solution is given
as  
$$
S = \left(\matrix{-\frac{1}{2} i & \frac{1}{2} i & \frac{1}{4} & 
-\frac{1}{4} \vspace*{0.2cm} \cr 
\frac{1}{2} i & - \frac{1}{2} i & \frac{1}{4} & 
-\frac{1}{4} \vspace*{0.2cm} \cr
1 & 1 & \frac{1}{2} & \frac{1}{2} \vspace*{0.2cm} \cr
-1 & -1 & \frac{1}{2} & \frac{1}{2}}\right) 
\hspace*{0.5cm}
T= \left(\matrix{e^{2\pi i / 12} & 0 & 0 & 0 \cr
0 & e^{2\pi i / 12} & 0 & 0 \cr
0 & 0 & e^{- \pi i / 12} & 0 \cr
0 & 0 & 0 & e^{11 \pi i / 12} }\right) \,.
$$
It is intriguing that a formal application of Verlinde's
formula \cite{verlinde} leads to fusion rule coefficients which are
positive integers.  These do not reproduce the fusion rules we 
have calculated. This is not surprising, as, for example, this set of 
representations does not contain the vacuum representation, {\it i.e.}
a representation which has trivial fusion rules. Even more
drastically, the fusion matrix corresponding to the representation 
${\cal V}_{-1/8}$ which, in the above basis, is given as
$$
N_{-1/8} = \left(\matrix{0 & 0 & 1 & 0 \cr 
0 & 0 & 0 & 1 \cr
2 & 2 & 0 & 0\cr
2 & 2 & 0 & 0} \right)\,,
$$ 
is not diagonalisable, and the same is true for the matrix
corresponding to ${\cal V}_{3/8}$. However, a slight modification of
Verlinde's observation still holds: the above $S$ matrix transforms
the fusion matrices into block diagonal form, where the blocks correspond
to the two one-dimensional and the one two-dimensional representation
of the fusion algebra.
\newpage

\section{The local triplet theory}
\renewcommand{\theequation}{4.\arabic{equation}}
\setcounter{equation}{0}

As we have seen in the previous section, the $c=-2$ triplet theory is
a chiral conformal field theory that is rational in the sense that it
only possesses finitely many indecomposable representations that close
under fusion. However, as we have also seen, some of the correlation 
functions contain logarithmic branch cuts, and the modular properties
of the representations are quite unusual. It is therefore an
interesting question whether one can define a consistent {\it local}
theory whose two chiral halves are the chiral triplet theory. As we
shall see, this is indeed possible (and the corresponding partition
function is indeed modular invariant), but a number of novel features
arise \cite{GKau98}. 

As in the usual case, we shall begin by trying to construct a local
theory whose space of states is the direct sum of tensor products of
the various chiral representations. As we have argued above, the
natural set of representations to consider consists of the two
irreducible representations ${\cal V}_{-1/8}$ and ${\cal V}_{3/8}$
together with the two indecomposable representations ${\cal R}_{0}$
and ${\cal R}_1$. Thus our ansatz for the total space of states is 
\be
\H^{ansatz} = \left({\cal V}_{-1/8}\otimes \bar{\cal V}_{-1/8}\right)
\bigoplus
\left({\cal V}_{3/8}\otimes \bar{\cal V}_{3/8}\right)
\bigoplus 
\left({\cal R}_0\otimes\bar{\cal R}_0\right)
\bigoplus \left({\cal R}_1\otimes\bar{\cal R}_1\right)\,.
\ee
As we shall see momentarily, $\H^{ansatz}$ is not quite
correct. One reason for this is actually easy to understand: the
M\"obius symmetry implies that the 2-point function satisfies 
\begin{displaymath}
  z\partial_z \langle\phi_1(z,\bar z)\phi_2(0,0)\rangle +
  \langle L_0\phi_1(z,\bar z)\phi_2(0,0)\rangle +
  \langle\phi_1(z,\bar z)L_0\phi_2(0,0)\rangle = 0 \,,
\end{displaymath}
and similarly for the barred coordinates. Integrating the difference
of these differential equations along a circle around the origin, 
we then get
\begin{displaymath}
  \langle\phi_1(\e^{-2\pi i}z,\e^{2\pi i}\bar z)\phi_2(0,0)\rangle =
  \e^{2\pi i(h_1-\bar h_1+h_2-\bar h_2)}
  \langle \e^{2\pi i S}\phi_1(z,\bar z)\e^{2\pi i S}\phi_2(0,0)\rangle\,,
\end{displaymath}
where $(h_j,\bar h_j)$ are the left and right conformal weights of the
states $\phi_j$ and $S = L_0^{(\mathrm{n})} - \bar L_0^{(\mathrm{n})}$
is the nilpotent part of $L_0 - \bar L_0$. The conditions for the
two-point function to be local are thus
\begin{displaymath}
  h_1-\bar h_1+h_2-\bar h_2 \in \Zop\,, \qquad
  \langle S^n \phi_1(z,\bar z) S^m \phi_2(0,0) \rangle = 0  \qquad  
  \forall
  n,m\in\Zop_{\geq0}, \, m+n>0 \,.
\end{displaymath}
This has to hold for any combination of $\phi_1$ and $\phi_2$.  Since
every $N$-point function involving $\phi_1$, say, can be expanded in
terms of such two-point functions (by defining $\phi_2$ to be a
suitable contour integral of the product of the remaining $N-1$
fields), it follows that we have to have
\begin{displaymath}
  h-\bar h \in \Zop\,, \qquad
  S \phi = 0  \,,
\end{displaymath}
where $(h,\bar h)$ are the conformal weights of any non-chiral  
field $\phi$. The first condition is well-known, but the second only
arises for theories for which $L_0$ (and $\bar L_0$) has a nilpotent
part. It is therefore new, and leads to a non-trivial
constraint. Consider for example the local fields associated to  
${\cal R}_0\otimes \bar{\cal R}_0$. The `ground states' are 
\be
\omega\otimes\bar\omega\qquad  
\begin{array}{l}
\omega\otimes\bar\Omega \\
\Omega\otimes\bar\omega
\end{array} \qquad 
\Omega\otimes\bar\Omega \,.
\ee
Now 
\begin{eqnarray}
S (\omega\otimes\bar\omega) & = & (\Omega\otimes\bar\omega) -
(\omega\otimes\bar\Omega) \\
S (\Omega\otimes\bar\omega) & = & (\Omega\otimes\bar\Omega) \,.
\end{eqnarray}
Thus $S\phi=0$ is not satisfied on all of 
${\cal R}_0\otimes \bar{\cal R}_0$. 

The natural resolution of this problem is to consider instead of 
${\cal R}_0\otimes \bar{\cal R}_0$ a suitable quotient space
\be
{\cal R}_{0\bar0} \equiv 
\left({\cal R}_0\otimes\bar{\cal R}_0\right) /   {\cal N}_{0\bar0}\,,
\ee
where ${\cal N}_{0\bar0}$ is chosen so that 
$S|_{{\cal R}_{0\bar0}}=0$. To determine this quotient space, we
observe that $S$ commutes with the action of both chiral
algebras. This implies that every state that is in the image space of
$S$, is in the subrepresentation generated from 
$S(\omega \otimes\bar\omega)$.\footnote{Here and in the following
every `representation' is a  representation of $\A\otimes\bar\A$,
where $\A$ and $\bar\A$ are the left- and right-moving triplet
algebra, respectively.} Thus the `minimal' choice for 
${\cal N}_{0\bar0}$ is to take it to be the subrepresentation of  
${\cal R}_0\otimes \bar{\cal R}_0$ generated from 
$S(\omega\otimes\bar\omega)=(\omega\otimes\bar\Omega)
-(\Omega\otimes\bar\omega)$. In particular, we note that 
${\cal N}_{0\bar0}$ then also contains the state 
$L_0 S (\omega\otimes\bar\omega) = (\Omega\otimes\bar\Omega)$.

By construction, $S|_{{\cal R}_{0\bar0}}=0$, and 
${\cal R}_{0\bar0}$ is still a representation of
$\A\otimes\bar\A$. The fact that ${\cal R}_{0\bar0}$ is
non-trivial is a consequence of the property of ${\cal R}_0$ and 
$\bar{\cal R}_0$ not to be irreducible as representations of $\A$ and 
$\bar\A$, respectively.

The representation ${\cal R}_{0\bar0}$ is still {\it not} an
irreducible representation of $\A\otimes\bar\A$: its space of ground
states is two-dimensional and can be taken to be spanned by  
\be
\bomega\equiv \omega\otimes\bar\omega \qquad \hbox{and}  \qquad 
\bOmega\equiv \half \left( \omega\otimes\bar\Omega 
              + \Omega\otimes\bar\omega\right)\,.
\ee
We then find that 
\begin{eqnarray}
L_0 \, \bomega = \Omega\otimes\bar\omega & = & 
\half \left( \omega\otimes\bar\Omega 
              + \Omega\otimes\bar\omega\right)
+ \half \left( (\Omega\otimes\bar\omega)
              - (\omega\otimes\bar\Omega)\right) \nonumber \\
& \simeq & \half \left( \omega\otimes\bar\Omega 
              + \Omega\otimes\bar\omega\right) = \bOmega\,,
\end{eqnarray}
and similarly for $\bar L_0\, \bomega = \bOmega$. 

\noindent For ${\cal R}_1\otimes\bar{\cal R}_1$, the situation is
analogous, and the relevant quotient space is 
\be
{\cal R}_{1\bar1} = \left({\cal R}_1\otimes\bar{\cal R}_1\right) /
{\cal N}_{1\bar1}\,,
\ee
where ${\cal N}_{1\bar1}$ is the subrepresentation generated from 
$\phi^\alpha\otimes\bar\psi^{\bar\alpha}
-\psi^\alpha\otimes\bar\phi^{\bar\alpha}$. The structure of the
resulting representations is therefore 
\begin{displaymath}
  \begin{picture}(184,210)(-92,-50)
    \put(2,40){\vbox to 0pt
        {\vss\hbox to 0pt{\hss$\bullet$\hss}\vss}}
    \put(2,90){\vbox to 0pt
        {\vss\hbox to 0pt{\hss$\bullet$\hss}\vss}}
    \put(62,0){\vbox to 0pt
        {\vss\hbox to 0pt{\hss$\bullet$\hss}\vss}}
    \put(56,4){\vector(-3,2){48}}
    \put(58,6){\vector(-2,3){52}}
    \put(-2,40){\vbox to 0pt
        {\vss\hbox to 0pt{\hss$\bullet$\hss}\vss}}
    \put(-2,90){\vbox to 0pt
        {\vss\hbox to 0pt{\hss$\bullet$\hss}\vss}}
    \put(-60,132){\vbox to 0pt
        {\vss\hbox to 0pt{\hss$\bullet$\hss}\vss}}
    \put(-60,128){\vbox to 0pt
        {\vss\hbox to 0pt{\hss$\bullet$\hss}\vss}}
    \put(-64,132){\vbox to 0pt
        {\vss\hbox to 0pt{\hss$\bullet$\hss}\vss}}
    \put(-64,128){\vbox to 0pt
        {\vss\hbox to 0pt{\hss$\bullet$\hss}\vss}}
    \put(-62,0){\vbox to 0pt
        {\vss\hbox to 0pt{\hss$\bullet$\hss}\vss}}
    \put(-8,36){\vector(-3,-2){48}}
    \put(-6,84){\vector(-2,-3){52}}
    \put(-8,94){\vector(-3,2){48}}
    \put(-6,46){\vector(-2,3){52}}
    \put(55,0){\vector(-1,0){110}}
    \put(-67,-15){\hbox to 0pt{\hss$\bOmega$}}
    \put(67,-15){\hbox to 0pt{$\bomega$\hss}}
    \put(-67,135){\hbox to 0pt{\hss$X^j_{-1}\bar  
X^{\bar\jmath}_{-1}\bomega$}}
    \put(47,40){\hbox to 0pt{$X^j_{-1}\bomega$\hss}}
    \put(17,90){\hbox to 0pt{$\bar X^{\bar\jmath}_{-1}\bomega$\hss}}
    \put(0,-40){\hbox to 0pt{\hss${\cal R}_{0\bar0}$\hss}}
  \end{picture}
\qquad
  \begin{picture}(184,210)(-92,-50)
    \put(2,40){\vbox to 0pt
        {\vss\hbox to 0pt{\hss$\bullet$\hss}\vss}}
    \put(2,90){\vbox to 0pt
        {\vss\hbox to 0pt{\hss$\bullet$\hss}\vss}}
    \put(62,130){\vbox to 0pt
        {\vss\hbox to 0pt{\hss$\bullet$\hss}\vss}}
    \put(56,126){\vector(-3,-2){48}}
    \put(58,124){\vector(-2,-3){52}}
    \put(-2,40){\vbox to 0pt
        {\vss\hbox to 0pt{\hss$\bullet$\hss}\vss}}
    \put(-2,90){\vbox to 0pt
        {\vss\hbox to 0pt{\hss$\bullet$\hss}\vss}}
    \put(-62,130){\vbox to 0pt
        {\vss\hbox to 0pt{\hss$\bullet$\hss}\vss}}
    \put(-60,2){\vbox to 0pt
        {\vss\hbox to 0pt{\hss$\bullet$\hss}\vss}}
    \put(-60,-2){\vbox to 0pt
        {\vss\hbox to 0pt{\hss$\bullet$\hss}\vss}}
    \put(-64,2){\vbox to 0pt
        {\vss\hbox to 0pt{\hss$\bullet$\hss}\vss}}
    \put(-64,-2){\vbox to 0pt
        {\vss\hbox to 0pt{\hss$\bullet$\hss}\vss}}
    \put(-8,36){\vector(-3,-2){48}}
    \put(-6,84){\vector(-2,-3){52}}
    \put(-8,94){\vector(-3,2){48}}
    \put(-6,46){\vector(-2,3){52}}
    \put(55,130){\vector(-1,0){110}}
    \put(-67,-15){\hbox to 0pt{\hss$\bxi^{\alpha\bar\alpha}$}}
    \put(-67,135){\hbox to 0pt{\hss$\bpsi^{\alpha\bar\alpha}$}}
    \put(67,135){\hbox to 0pt{$\bphi^{\alpha\bar\alpha}$\hss}}
    \put(17,40){\hbox to 0pt{$\brho^{\alpha\bar\alpha}$\hss}}
    \put(47,90){\hbox to 0pt{$\bar\brho^{\alpha\bar\alpha}$\hss}}
    \put(0,-40){\hbox to 0pt{\hss${\cal R}_{1\bar1}$\hss}}
  \end{picture}
  \begin{picture}(40,210)(-10,-50)
    \put(0,0){(0,0)}
    \put(0,40){(1,0)}
    \put(0,90){(0,1)}
    \put(0,130){(1,1)}
    \put(-5,-40){weight}
  \end{picture}
\end{displaymath}
Here $\bphi^{\alpha\bar\alpha}$ is the equivalence class in 
${\cal R}_{1\bar1}$ with representative 
$(\phi^\alpha \otimes \bar\phi^{\bar\alpha})$. 

The above constraints are those that arise from the requirement that
the $2$-point functions are local. The analysis of the locality of the
higher correlation functions is actually quite complicated (see
\cite{GKau98} for a detailed description), but the final constraint is
rather simple: in order to obtain local correlation functions the
states at grade $(0,1)$ and the states at $(1,1)$ in the two
representations ${\cal R}_{0\bar0}$ and ${\cal R}_{1\bar1}$ have 
to be identified
\begin{center}
  \begin{picture}(184,210)(-92,-50)
    \put(2,40){\vbox to 0pt
        {\vss\hbox to 0pt{\hss$\bullet$\hss}\vss}}
    \put(2,90){\vbox to 0pt
        {\vss\hbox to 0pt{\hss$\bullet$\hss}\vss}}
    \put(62,130){\vbox to 0pt
        {\vss\hbox to 0pt{\hss$\bullet$\hss}\vss}}
    \put(62,0){\vbox to 0pt
        {\vss\hbox to 0pt{\hss$\bullet$\hss}\vss}}
    \put(56,4){\vector(-3,2){48}}
    \put(58,6){\vector(-2,3){52}}
    \put(56,126){\vector(-3,-2){48}}
    \put(58,124){\vector(-2,-3){52}}
    \put(-2,40){\vbox to 0pt
        {\vss\hbox to 0pt{\hss$\bullet$\hss}\vss}}
    \put(-2,90){\vbox to 0pt
        {\vss\hbox to 0pt{\hss$\bullet$\hss}\vss}}
    \put(-62,130){\vbox to 0pt
        {\vss\hbox to 0pt{\hss$\bullet$\hss}\vss}}
    \put(-62,0){\vbox to 0pt
        {\vss\hbox to 0pt{\hss$\bullet$\hss}\vss}}
    \put(-8,36){\vector(-3,-2){48}}
    \put(-6,84){\vector(-2,-3){52}}
    \put(-8,94){\vector(-3,2){48}}
    \put(-6,46){\vector(-2,3){52}}
    \put(55,0){\vector(-1,0){110}}
    \put(55,130){\vector(-1,0){110}}
    \put(-67,-15){\hbox to 0pt{\hss$\bOmega$}}
    \put(67,-15){\hbox to 0pt{$\bomega$\hss}}
    \put(-67,135){\hbox to 0pt{\hss$\bpsi^{\alpha\bar\alpha}$}}
    \put(67,135){\hbox to 0pt{$\bphi^{\alpha\bar\alpha}$\hss}}
    \put(47,40){\hbox to 0pt{$\brho^{\alpha\bar\alpha}$\hss}}
    \put(47,90){\hbox to 0pt{$\bar\brho^{\alpha\bar\alpha}$\hss}}
    \put(0,-40){\hbox to 0pt{\hss${\cal R}$\hss}}
  \end{picture}
\qquad
  \begin{picture}(40,210)(-10,-50)
    \put(0,0){(0,0)}
    \put(0,40){(1,0)}
    \put(0,90){(0,1)}
    \put(0,130){(1,1)}
    \put(-5,-40){weight}
  \end{picture}
\end{center}
The resulting representation ${\cal R}$ does not have one cyclic
state, but instead is generated from a state $\bomega$ of weight
$(0,0)$ and the four states $\bphi^{\alpha\bar\alpha}$ of weight
$(1,1)$. The non-trivial defining relations are
\begin{displaymath}
  \begin{array}{rcl@{\qquad}rcl}
    L_0 \bomega &=& \bOmega, & W^a_0 \bomega &=& 0, \\ 
    L_0 \bOmega &=& 0, & W^a_0 \bOmega &=& 0, \\ 
    L_{-1} \bomega &=& \tilde\Theta_{\bar\alpha\alpha}
    \brho^{\alpha\bar\alpha}, & 
    W^a_{-1} \bomega &=& t^{a\alpha}_\beta \tilde\Theta_{\bar\alpha\alpha}
    \brho^{\beta\bar\alpha}, \\ [\bigskipamount]
    L_0 \brho^{\alpha\bar\alpha} &=& 
    \brho^{\alpha\bar\alpha}, & 
    W^a_0 \brho^{\alpha\bar\alpha} &=& 
    2t^{a\alpha}_\beta \brho^{\beta\bar\alpha}, \\ 
    L_1 \brho^{\alpha\bar\alpha} &=& 
    \Theta^{\alpha\bar\alpha} \, \bOmega, & 
    W^a_1 \brho^{\alpha\bar\alpha} &=& 
    t^{a\alpha}_\beta \Theta^{\beta\bar\alpha} \, \bOmega, \\ 
    L_{-1} \bar\brho^{\alpha\bar\alpha} &=& 
    \bpsi^{\alpha\bar\alpha}, & 
    W^a_{-1} \bar\brho^{\alpha\bar\alpha} &=& 
    t^{a\alpha}_\beta \bpsi^{\beta\bar\alpha}, \\[\bigskipamount]
    L_0 \bphi^{\alpha\bar\alpha} &=& 
    \bphi^{\alpha\bar\alpha} + \bpsi^{\alpha\bar\alpha}, & 
    W^a_0 \bphi^{\alpha\bar\alpha} &=& 
    2 t^{a\alpha}_\beta \bphi^{\beta\bar\alpha}, \\ 
    L_0 \bpsi^{\alpha\bar\alpha} &=& 
    \bpsi^{\alpha\bar\alpha}, & 
    W^a_0 \bpsi^{\alpha\bar\alpha} &=& 
    2 t^{a\alpha}_\beta \bpsi^{\beta\bar\alpha}, \\ 
    L_1 \bphi^{\alpha\bar\alpha} &=& 
    - \bar\brho^{\alpha\bar\alpha}, & 
    W^a_1 \bphi^{\alpha\bar\alpha} &=& 
    -t^{a\alpha}_\beta \bar\brho^{\beta\bar\alpha}, 
  \end{array}
\end{displaymath}
together with their anti-chiral counterparts. Here $t^{a\alpha}_\beta$
are the matrix elements of the spin $\half$ representation of $su(2)$,
and $\Theta^{\alpha\bar\beta}$ and $\tilde\Theta_{\bar\alpha\beta}$
are tensors that are described in detail in \cite{GKau98}.

In addition to the sates in ${\cal R}$ we also have the states
that are associated to the tensor products of the irreducible
representations,  
\be
{\cal  V}_{-1/8,-1/8} = 
\left({\cal V}_{-1/8} \otimes \bar{\cal V}_{-1/8}\right)
\qquad \hbox{and} \qquad 
{\cal V}_{3/8,3/8} = 
\left({\cal V}_{3/8} \otimes \bar{\cal V}_{3/8}\right) \,.
\ee
As was shown in \cite{GKau98} it is then possible to `solve the
bootstrap' for this space of states, \ie\ to find operator product
coefficients for all possible operator products involving these states
so that the amplitudes of the resulting theory are local.

\subsection{The partition function}

In the above we have analysed the locality of the various
amplitudes. As we have mentioned, these constraints were already
sufficient to determine the structure of the resulting representation
${\cal R}$ uniquely. Given the structure of the various local
representations, we can ask whether the partition function of the
whole theory is modular invariant. 

First of all, the characters of the non-chiral irreducible
representations are simply the product of a left and right chiral
character 
\begin{eqnarray*}
  \chi_{{\cal V}_{-1/8,-1/8}}(\tau) &=& 
  \chi_{{\cal V}_{-1/8}}(\tau) \bar\chi_{{\cal V}_{-1/8}}(\bar\tau) = 
  \left| \eta(\tau)^{-1} \theta_{0,2}(\tau) \right|^2 \,, 
  \\*
  \chi_{{\cal V}_{3/8,3/8}}(\tau) &=& 
  \chi_{{\cal V}_{3/8}}(\tau) \bar\chi_{{\cal V}_{3/8}}(\bar\tau) = 
  \left| \eta(\tau)^{-1} \theta_{2,2}(\tau) \right|^2\,, 
\end{eqnarray*}
where we are using the same notation as in section 3.7. To determine
the character of the reducible representation ${\cal R}$, let us
recall that each vertex in the diagrammatical representation of 
${\cal R}$ corresponds to an irreducible representation of the left
and right triplet algebra. Putting the different contributions
together we find 
\begin{eqnarray*}
  \chi_{{\cal R}}(\tau) &=& 
  2 \chi_{{\cal R}_0}(\tau) \bar\chi_{{\cal R}_0}(\bar\tau) + 
  2 \chi_{{\cal R}_1}(\tau) \bar\chi_{{\cal R}_0}(\bar\tau) + 
  2 \chi_{{\cal R}_0}(\tau) \bar\chi_{{\cal R}_1}(\bar\tau) + 
  2 \chi_{{\cal R}_1}(\tau) \bar\chi_{{\cal R}_1}(\bar\tau)
  \\*
  &=&
  2 \left| \chi_{{\cal R}_0}(\tau) + \chi_{{\cal R}_1}(\tau) \right|^2
  \\*
  &=&
  2 \left| \eta(\tau)^{-1} \theta_{1,2}(\tau) \right|^2 \,.
\end{eqnarray*}
The partition function of the full theory is thus
\begin{displaymath}
  Z = 
  \chi_{{\cal V}_{-1/8,-1/8}}(\tau) + 
  \chi_{{\cal V}_{3/8,3/8}}(\tau) + 
  \chi_{{\cal R}}(\tau)
  =
  \left|\eta(\tau)\right|^{-2} 
  \sum_{k=0}^3 \left|\theta_{k,2}(\tau)\right|^2 \,,
\end{displaymath}
and this is indeed modular invariant as follows directly from
(\ref{thetatrans}) and (\ref{etatrans}). Actually, the above partition 
function is the same as that of a free boson compactified on a circle
of radius $\sqrt{2}$ \cite{Ginsparg}. However, in our case the
partition function is not simply the sum of products of left- and
right- chiral representations of the chiral algebra as the non-chiral
representation ${\cal R}$ is {\it not} the tensor product of a left-
and right-chiral representation, but only a quotient thereof. As we
have explained before, this follows directly from locality.

\subsection{Symplectic fermions}

Up to now we have constructed the local theory abstractly, starting 
from the chiral representations of the triplet algebra and determining
the constraints that arise from locality. In this section we want to
show that the resulting conformal field theory is actually the same as
the bosonic sector of a free field theory of `symplectic' fermions.
Here we shall only summarise the essential features of this fermion
model; further details may be found in \cite{Kausch98} and
\cite{Kau95}.   

The chiral algebra of the symplectic fermion model is generated by a
two-component fermion field $\chi^\alpha$ of conformal weight
one whose anti-commutator is given by
\begin{displaymath}
  \{ \chi^\alpha_m, \chi^\beta_n \} = m\, d^{\alpha\beta}\,
  \delta_{m+n}\,, 
\end{displaymath}
where $d^{\alpha\beta}$ is an anti-symmetric tensor with components 
$d^{\pm\mp} = \pm1$. This algebra has a unique irreducible highest
weight representation generated from a highest weight state $\Omega$
satisfying $\chi^\alpha_m \Omega = 0$ for $m\geq0$; we may call this
representation the vacuum representation. It contains the triplet
$W$-algebra since 
\begin{eqnarray*}
  L_{-2} \Omega &=& 
  \frac12 d_{\alpha\beta} \chi^\alpha_{-1} \chi^\beta_{-1} \Omega \,, 
  \\
  W^a_{-3} \Omega &=& 
  t^a_{\alpha\beta} \chi^\alpha_{-2} \chi^\beta_{-1} \Omega  \,,
\end{eqnarray*}
satisfy the triplet algebra \cite{Kausch91,Kau95}. Here
$d_{\alpha\beta}$ is the inverse metric to $d^{\alpha\beta}$, and
$t^a_{\alpha\beta}$ are the matrix elements of the spin $\half$
representation of $su(2)$. 

Let us consider the maximal highest weight representation of this
chiral algebra that contains the vacuum representation. It is
generated by the negative modes $\chi^\alpha_m, m<0$ from a four
dimensional space of ground states. This space is spanned by two
bosonic states $\Omega$ and $\omega$, and two fermionic states,
$\theta^\alpha$, and the action of the zero-modes $\chi^\alpha_0$ is 
given as  
\begin{eqnarray*}
  \chi^\alpha_0 \omega &=& -\theta^\alpha\,, 
  \\
  \chi^\alpha_0 \theta^\beta &=& d^{\alpha\beta} \Omega\,, 
  \\
  L_0 \omega &=& \Omega\,. 
\end{eqnarray*}
Imposing the locality constraints as above, the corresponding
non-chiral representation is then generated by the negative modes
$\chi^\alpha_m, \bar\chi^{\bar\alpha}_m, m<0$ from the ground space
representation   
\begin{displaymath}
  \begin{array}{rcl@{\qquad}rcl}
    \chi^\alpha_0 \bomega &=& -\btheta^\alpha\,, &
    \bar\chi^{\bar\alpha}_0 \bomega &=& -\bar\btheta^{\bar\alpha}\,, 
    \\
    \chi^\alpha_0 \btheta^\beta &=& d^{\alpha\beta} \bOmega\,, &
    \bar\chi^{\bar\alpha}_0 \bar\btheta^{\bar\beta} &=& 
    d^{\bar\alpha\bar\beta} \bOmega\,, 
    \\
    \chi^\alpha_0 \bar\btheta^{\bar\alpha} &=&
    \Theta^{\alpha\bar\alpha} \bOmega\,, &
    \bar\chi^{\bar\alpha}_0 \btheta^\alpha &=&
    -\Theta^{\alpha\bar\alpha} \bOmega\,,
  \end{array}
\end{displaymath}
where $\Theta^{\alpha\bar\alpha}$ is given as before. The space of
ground states contains two bosonic states, $\bOmega$ and $\bomega$,
and two fermionic states since the four fermionic states
$\btheta^\alpha$ and $\bar\btheta^{\bar\alpha}$ are related as  
\begin{displaymath}
  \btheta^\alpha = 
  \Theta^{\alpha\bar\alpha} d_{\bar\alpha\bar\beta}
  \bar\btheta^{\bar\beta}\,, \qquad
  \bar\btheta^{\bar\alpha} = 
  - \Theta^{\alpha\bar\alpha} d_{\alpha\beta} \btheta^\beta\,. 
\end{displaymath}
One can show (see \cite{Kausch98} for further details) that the 
bosonic sector of this representation is isomorphic to the
representation ${\cal R}$. For example, the higher level states of
${\cal R}$ can be expressed as fermionic descendents as 
\begin{eqnarray*}
  \brho^{\alpha\bar\alpha} &=& \chi^\alpha_{-1}
  \bar\btheta^{\bar\alpha}\,, \\
  \bar\brho^{\alpha\bar\alpha} &=& -\bar\chi^{\bar\alpha}_{-1}
  \btheta^\alpha\,, \\
  \bpsi^{\alpha\bar\alpha} &=& 
  \chi^\alpha_{-1}\bar\chi^{\bar\alpha}_{-1}\bOmega\,, \\
  \bphi^{\alpha\bar\alpha} &=& 
  \chi^\alpha_{-1}\bar\chi^{\bar\alpha}_{-1}\bomega\,.
\end{eqnarray*}
We should mention that, as the bosonic sector of the symplectic
fermion representation, the representation ${\cal R}$ is 
generated from a single cyclic state $\bomega$.

The other two representations of the triplet model, the irreducible
representations ${\cal  V}_{-1/8,-1/8}$ and ${\cal V}_{3/8,3/8}$, also
have an interpretation in terms of the symplectic fermion theory: they
correspond to the bosonic sector of the (unique)
$\Zop_2$-twisted representation. In this sector, the fermions
are half-integrally moded, but all bosonic operators (including the
triplet algebra generators that are bilinear in the fermions) are
still integrally moded. The ground state of the twisted sector,
$\bmu$, has conformal weight $h=\bar{h}=-1/8$ and satisfies
\begin{displaymath}
  \chi^\alpha_{r}\bmu = \bar\chi^{\bar\alpha}_{r}\bmu = 0\,, 
  \qquad\textrm{for $r>0$}\,, 
\end{displaymath}
while
\begin{displaymath}
  \bnu^{\alpha\bar\alpha} = 
  \chi^\alpha_{-\frac12}\bar\chi^{\bar\alpha}_{\frac12}\bmu\,. 
\end{displaymath}
With respect to the symplectic fermions, $\bomega$ and $\bmu$ are
cyclic states, and all amplitudes can be reduced to those involving
$\bmu$ and the four ground states of the vacuum representation. This
can be done using the (fermionic) comultiplication formula and its
twisted analogue \cite{Gab97}. The amplitudes involving the ground
states can then be determined by solving differential equations.   
One can check that the amplitudes of the symplectic fermion theory
reproduce the amplitude constructed above \cite{GKau98}, and
therefore that the two theories agree. 
\bigskip

This concludes our analysis of the triplet theory at $c=-2$. In the
next section we want to discuss another theory that also exhibits
logarithmic behaviour. 

\newpage

\section{Logarithmic representations at fractional level}
\renewcommand{\theequation}{5.\arabic{equation}}
\setcounter{equation}{0}

One of the best understood conformal field theories is the WZW model
that can be defined for any (simple compact) group \cite{wzw}. If the
so-called level is chosen to be a positive integer, the theory is
unitary and rational, and in fact these models are the paradigm for
rational conformal field theories. The fusion rules are well known
\cite{gepwit,kac,walton,furlan}, and they can be obtained, via the 
Verlinde formula \cite{verlinde}, from the modular transformation
properties of the characters.  

{} From a Lagrangian point of view, the model is only well defined if
the level is integer, but the corresponding vertex operator algebra
(or the meromorphic conformal field theory in the sense of
\cite{GabGod}) can also be constructed even if this is not the
case. Furthermore, it was realised some time ago that there exists a
preferred set of admissible (fractional) levels for which the
characters corresponding to the `admissible' representations have
simple modular properties \cite{kacwac}. This suggests that these
admissible level WZW models define `almost' rational conformal field
theories. However, there are clearly some subtleties since the fusion
rule coefficients that can be obtained by the application of the
Verlinde formula are integers, but not all positive integers 
\cite{kohsorba,matwal}.
\smallskip

The fusion rules of WZW models at admissible fractional level have
been studied quite extensively over the years. In particular, the
simplest case of $su(2)$ at fractional level has been analysed in
detail \cite{berfel,panda,ay,dot,fm,andreev,dlm,pry,fgp,gpw}  
(for a good review about the various results see in particular
\cite{gpw}). All of these fusion rule calculations essentially
determine the possible couplings of three representations. More
precisely, given two representations, the calculations determine
whether a given third representation can be contained in the fusion
product of the former two.   

Two different sets of `fusion rules' have been proposed in the
literature: the fusion rules of Bernard and Felder \cite{berfel}
whose calculations have been reproduced in \cite{dot,dlm}, and the
fusion rules of Awata and Yamada \cite{ay} whose results have been
recovered in \cite{fm,andreev,pry,fgp,gpw}. The two calculations
differ essentially by what class of representations is considered: in
Bernard \& Felder only admissible representations that are highest
weight with respect to the whole affine algebra (and a fixed choice of
a Borel subalgebra) are considered, while in  Awata \& Yamada also
representations that are highest weight with respect to an arbitrary
Borel subalgebra were analysed. As a consequence, the fusion rules of
Awata \& Yamada `contain' the fusion rules of Bernard \& Felder. In
addition, the fusion analysis of Awata \& Yamada is based on a
generalised notion of fusion for which the correlation functions are
also regarded as functions of the variables that characterise the
different Borel subalgebras. (This was clarified in
\cite{fm,fmone}.) This modified notion of fusion is physically not 
very well motivated, and it falls outside the usual framework of
vertex operator algebras; in the following we shall therefore consider
the usual definition of fusion. As we shall see, the corresponding
fusion rules are commutative and associative, and they close on a
certain set of representations that we shall describe in detail
\cite{Gab01}.  

In deriving the `fusion rules' from the calculations in
\cite{berfel,dot,dlm}, it is always assumed implicitly that the
actual fusion product is a direct sum of representations of the kind
that are considered. (This is to say, there are no additional fusion
channels that one has overlooked by restricting oneself to the class
of representations in question.) In particular, it has been assumed
that the fusion rules `close' on (conformal) highest weight
representations (since these are the only representations that were  
considered). However, as we shall explain in some detail in the
following, this is not true in general. In fact, the fusion product of
two highest weight representations (with respect to the affine
algebra) contains sometimes a representation whose $L_0$ spectrum is
not bounded from below \cite{Gab01}. (The representation has,
however, the property that $V_n(\psi)\chi=0$ for $n\geq N$, where $N$
depends on both $\psi$ and $\chi$; this is sufficient to guarantee
that the corresponding correlation functions do not have essential
singularities.) Furthermore, some of the representations that occur
are not completely decomposable, and indeed are logarithmic in the
sense that the action of $L_0$ is not diagonal \cite{Gab01}. For WZW 
models at level $k=0$ logarithmic representations have been discovered
before in \cite{CKT95,CKT96,Nichols,Nichols1}; however these models are
somewhat pathological (at $k=0$ the vacuum representation is
trivial), and the relevant logarithmic representations are quite
different from what will be described below.

\subsection{The example of $k=-4/3$}

In the following we shall concentrate on the simplest example, the
$su(2)$ model for $k=-4/3$. As we have mentioned before (in
section~3.2) the chiral algebra of this conformal field theory
contains the affine algebra $\hat{su}(2)$, whose modes satisfy 
the commutation relations
\begin{eqnarray}
{}[J^+_m,J^-_n] & = & 2 J^3_{m+n} + k m \delta_{m,-n} \nonumber \\
{}[J^3_m,J^\pm_n] & = & \pm J^\pm_{m+n} \nonumber \\
{}[J^3_m,J^3_n] & = & {k\over 2} m \delta_{m,-n}\,.
\end{eqnarray}
By virtue of the Sugawara construction, we can define Virasoro
generators as bilinears in the currents $J$
\be\label{sugawara}
L_n = {1\over (k+2)} \sum_l \left(
\half :J^+_l J^-_{n-l} : + \half  :J^-_l J^+_{n-l} : + 
: J^3_l J^3_{n-l} \right) \,,
\ee
where $: \cdot :$ denotes `normal ordering', \ie\ 
\be
: J^a_n J^b_m :  = \left\{
\begin{array}{ll}
J^a_n J^b_m & \hbox{if $n\leq m$} \\
J^b_m J^a_n & \hbox{if $n>m$.}
\end{array}
\right.
\ee
These Virasoro modes satisfy the commutation relations 
\begin{eqnarray}
{}[L_m,J^a_n] & = & - n J^a_{m+n} \nonumber \\
{}[L_m,L_n] & = & (m-n) L_{m+n} + {c \over 12} m (m^2-1)
\delta_{m,-n}\,, \nonumber
\end{eqnarray}
where $c$ is given in terms of the level $k$ as 
\be
c= {3 k \over (k+2)}\,.
\ee
In our case we have $k=-4/3$ and thus $c=-6$. 
\medskip

\noindent At $k=-4/3$, the vacuum representation has a null-vector at
level $3$ that is explicitly given by 
\be\label{vacnullsu2}
\Nu  = \left(J^3_{-3} 
      + {3\over 2} J^+_{-2} J^-_{-1} 
      - {3\over 2} J^+_{-1} J^-_{-2}
      + {9\over 2} J^3_{-1} J^+_{-1} J^-_{-1}
      + {9\over 2} J^3_{-1} J^3_{-1} J^3_{-1}
      - {9\over 2} J^3_{-2} J^3_{-1} \right) \Omega \,.
\ee
As we have explained before in section 3.2, every representation of
the conformal field theory must only contain states that are
annihilated by the zero mode of $\Nu$. In particular, if the
representation contains a highest weight state $\psi$, \ie\ a state
$\psi$ that is annihilated by all $J^a_n$ with $n>0$, we obtain the
condition 
\begin{eqnarray}
V_0(\N)\psi & = & \left[{9\over 2} \left( J^-_0 J^+_0 J^3_0 
+ J^3_0 J^3_0 J^3_0 + J^3_0 J^3_0 \right) + J^3_0 \right] \psi
\nonumber \\
& = & \left[ \left( {9\over 2} C + 1 \right) J^3_0 \right] \psi\,,
\label{condition1}
\end{eqnarray}
where we have denoted by $C$ the Casimir operator of the zero mode
$su(2)$ algebra, 
\be
C= {1\over 2} (J^+_0 J^-_0 + J^-_0 J^+_0) 
                 + J^3_0J^3_0\,.
\ee
Thus it follows from (\ref{condition1}) that the theory has two
types of highest weight representations:
\begin{list}{(\alph{enumi})}{\usecounter{enumi}}
\item the vacuum representation $\H_0$
\item any representation for which $C=-2/9$ on the highest weight
states.
\end{list}
It follows from (\ref{sugawara}) that on a highest weight state
$\psi$, 
\be
L_0 \, \psi = {1\over (k+2)} C \, \psi\,.
\ee
Thus the ground state of the vacuum representation has $h=0$, while
the ground states of the representations in (b) have $h=-1/3$. In
particular, it therefore follows from this analysis that $L_0$ acts
diagonally on every allowed highest weight representation! Thus it
would seem at first sight that this theory cannot contain any
logarithmic representations. As we shall explain further below, this 
is however not quite correct. 

It also follows from the above analysis that the theory is very far
from being `rational' since there exists a continuum of highest weight
representations of type (b). First of all, (b) includes the
representation $D^+_j$ with $j=-2/3$ and $j=-1/3$, where $D^+_j$ is
generated from a highest weight state $|j\rangle$, satisfying
\begin{eqnarray}
J^a_n |j\rangle & = & 0 \qquad \hbox{for any $a$ and $n>0$,} \nonumber \\
J^+_0 |j\rangle & = & 0 \nonumber \\
J^3_0 |j\rangle & = & j |j\rangle \,. \nonumber
\end{eqnarray}
In particular, the last two lines imply that the eigenvalue of $C$ on
$|j\rangle$ is equal to 
\be
C |j\rangle = j(j+1) |j\rangle \,.
\ee
Thus we have $C=-2/9$ provided that $j=-1/3$ or $j=-2/3$. Similarly, 
(b) also includes the representation $D^-_j$ with $j=2/3$ and $j=1/3$,
where $D^-j$ is generated from a highest weight state $|j\rangle$,
satisfying 
\begin{eqnarray}
J^a_n |j\rangle & = & 0 \qquad \hbox{for any $a$ and $n>0$,} \nonumber \\
J^-_0 |j\rangle & = & 0 \nonumber \\
J^3_0 |j\rangle & = & j |j\rangle \,. \nonumber
\end{eqnarray}
These four representations are special in that the $J^3_0$ spectrum on
the highest weight states is bounded from above (below) for $D^+_j$
($D^-_j$). They are precisely the {\it admissible} representations of
Kac \& Wakimoto \cite{kacwac}. 

However, there are also other representations that are compatible with
the vacuum representation and for which the spectrum of $J^3_0$ on the
highest weight states is not bounded: for every $t\in [0,1)$ with
$t\ne {1\over 3}, {2\over 3}$, there exists a representation $E_t$
that is generated from the highest weight states $|m\rangle$ with
$m-t\in\Zop$ for which we have
\begin{eqnarray}
J^a_n |m\rangle & = & 0 \qquad \hbox{for any $a$ and $n>0$} \nonumber \\
J^3_0 |m\rangle & = & m |m\rangle\nonumber \\
J^+_0 |m\rangle & = & |m+1\rangle \label{Ereps} \\
J^-_0 |m\rangle & = & \left(-{2\over 9} - m (m-1)\right)
          |m-1\rangle \,. \nonumber
\end{eqnarray}
It is easy to check that we have $C|m\rangle = -2/9 |m\rangle$, and
thus that $E_t$ defines an allowed highest weight representation of
the conformal field theory.

\subsection{An interesting symmetry}

As we shall show further below, the fusion rules of this theory
are actually quite complicated. A first indication of this fact may be
seen as follows. It is well known that the affine $\hat{su}(2)$
algebra has an outer automorphism given by 
\begin{eqnarray}
\pi_s(J^\pm_m) & = & J^\pm_{m\mp s} \nonumber \\
\pi_s(J^3_m) & = & J^3_m - {k\over 2} s\delta_{m,0}\,, \label{auto}
\end{eqnarray}
where $s\in\Zop$. The induced action on the Virasoro generators is
given by  
\be\label{virasym}
\pi_s(L_m) = L_m - s J^3_m + {1\over 4} k s^2\delta_{m,0}\,.
\ee 
If $s$ is even, the automorphism is inner in the sense that it can be 
obtained by the adjoint action of an element in the loop group of
$SU(2)$; on the other hand, if $s$ is odd, the automorphism can be
obtained by the adjoint action of a loop in $SO(3)$ that does not
define an element in the loop group of $SU(2)$ \cite{ps,mrgwzw}. 

As we have explained before in section 3.2, for positive integer $k$,
the allowed highest weight representations have highest weights that
transform in the $su(2)$ representation with 
$j=0,{1\over 2},\ldots,{k\over 2}$. In this case, the induced action
of the automorphism $\pi_1$ on the highest weight representations is
given by  
\be
\pi_1 : j \mapsto {k\over 2} - j \,.
\ee
In particular, $\pi_s$ with $s$ even maps each integrable positive
energy representations into itself; this simply reflects the fact that
every such representation gives rise to a representation of the full
loop group, and that the automorphism for $s$ even is inner (in the
sense described above).  

For positive integer $k$, the fusion rules respect this symmetry in
the sense that 
\be\label{symmetry}
\Bigl(\pi_s(\H_1) \otimes \pi_t(\H_2) \Bigr)_{\mathrm f} 
= \pi_{s+t} \Bigl(\left( \H_1 \otimes \H_2 \right)_{\mathrm f}
\Bigr)\,.
\ee
This seems to be quite a general property of `twist'-symmetries such
as (\ref{auto}) (see for example \cite{Gab97} for another example of
this type for the case of the ${\cal N}=2$ algebras); we shall
therefore assume in the following that the fusion rules also satisfy
this property in the fractional case. In any case, this is consistent
with what was found in \cite{Gab01}, where the fusion products were
analysed using the analogue of the fusion algorithm described in
section~2.2. 
\medskip

If we take (\ref{symmetry}) seriously it follows immediately that the
fusion rules of highest weight representations will not in general
define highest weight representations. Indeed, it is easy to check
that  
\begin{eqnarray}
\pi_1(\H_0) & = & D^-_{\twth} \\
\pi_{-1}(\H_0) & = & D^+_{-\twth} \\
\pi_1(D^+_{-\onth}) & = & D^-_{\onth}\,.
\end{eqnarray}
Thus we have 
\begin{eqnarray}
\Bigl(D^+_{-\twth} \otimes D^+_{-\twth} \Bigr)_{\mathrm f}
& = & 
\Bigl(\pi_{-1}(\H_0) \otimes \pi_{-1}(\H_0) \Bigr)_{\mathrm f}
\nonumber \\
& = & \pi_{-2} (\H_0)\,,
\end{eqnarray}
where we have used (\ref{symmetry}) in the last line (as well as the
fact that the fusion product of the vacuum representation with itself
is again the vacuum representation). It is now easy to check that
$\pi_{-2} (\H_0)$ is {\it not} a highest weight representation since
its $L_0$ spectrum is unbounded from below! However, it follows from
(\ref{virasym}) that  
\be
\pi_{-2} (L_0) -k = L_0 + 2 J^3_0  \,.
\ee
Furthermore, in the vacuum representation the spectrum of $|J^3_0|$ is
bounded by $L_0$, \ie\ 
\be
L_0 - |J^3_0| \geq 0 \,.
\ee
Thus we can conclude that we have 
\be
\pi_{-2}(L_0) - k \geq J^3_0 \,.
\ee
In particular, for each fixed eigenvalue of $J^3_0$, the $L_0$-spectrum
of $\pi_{-2} (\H_0)$ is bounded from below. This property is
sufficient to deduce, for example, that for any $\psi\in\H_0$ and any   
$\chi\in\pi_{-2}(\H_0)$, there exists a $N(\psi,\chi)$ such that
\be
V_n(\psi) \chi = 0 \qquad \hbox{for all $n\geq N(\psi,\chi)$.}
\ee
This in turn is sufficient to conclude that the correlation functions
involving states in $\pi_{-2}(\H_0)$ will not give rise to essential
singularities.

\subsection{Fusion rules}

As we have explained in the previous section, it follows from the
symmetry (\ref{symmetry}) that the fusion products of conventional
highest weight representations (such as $D^\pm_j$) will, in general,
not be highest weight, and will in fact involve representations whose
$L_0$ spectrum is unbounded from below. We shall now attempt to give a
detailed description of the full fusion rules. These fusion rules have
been determined using the analogue of the fusion algorithm described in
section 2.2. In the following we shall only sketch the results; 
more details can be found in \cite{Gab01}.

Because of the above remarks, all fusion products involving $\H_0$,
$D^\pm_{\mp \twth}$ are already determined by virtue of the symmetry
(\ref{symmetry}). If we start with the Kac-Wakimoto representations
$D^\pm_j$ the only fusion rule that is not determined by this symmetry
is then  
\be
\left(D^+_{-\onth} \otimes D^-_{\onth}\right)_{\mathrm f}
= \H_0 \oplus E_0 \,,
\ee
where $E_0$ is the (conformal) highest weight representation for which
the highest weight space is described by (\ref{Ereps}). Thus it
follows that the fusion rules do not close on the Kac-Wakimoto
representations (and their images under $\pi_s$). In order to analyse
the complete fusion rules we therefore have to determine the fusion
rules involving $E_0$. Again, because of (\ref{symmetry}) the only
fusion product that has to be calculated involves $D^-_{\onth}$, and
we find
\be
\left(D^{-}_{\onth} \otimes \H_E\right)_{\mathrm f} = \R_{\onth}\,.
\ee
The structure of the representation $\R_{\onth}$ is schematically
sketched in figure~1.
\begin{figure}[htb]
\epsfysize=6cm
\centerline{\epsffile{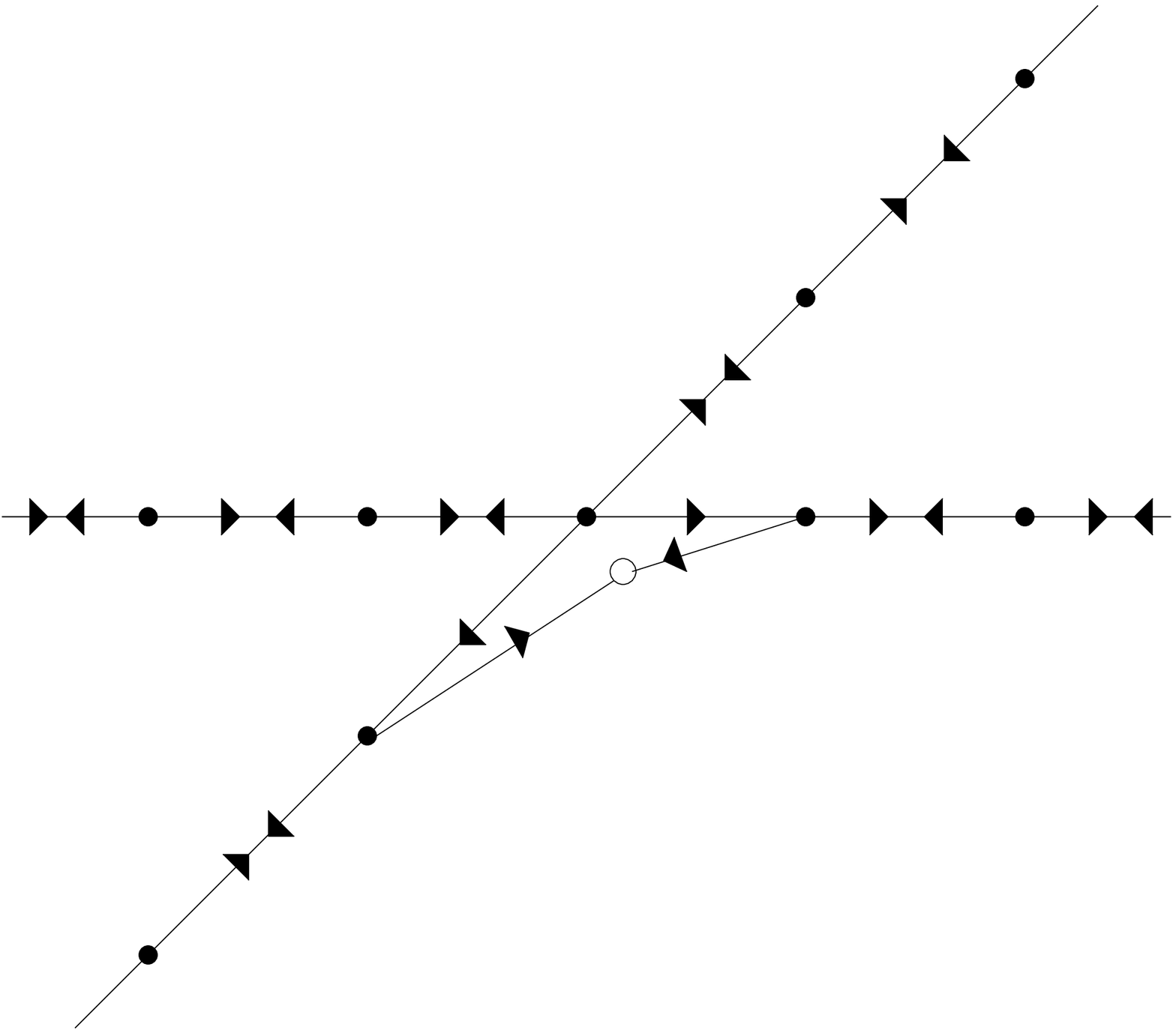}}
\caption{The representation $\R_{\onth}$.}
\end{figure}

\noindent Here we have arranged the circles representing the states
according to their charges, with the horizontal axis corresponding to
$J^3_0$, and the vertical axis to $L_0$. The horizontal array of
circles represent the states of the form $(m-\onth,-\onth)$ with
$m\in\Zop$, where the quantum numbers of a state $(m,h)$ are given as  
\begin{eqnarray}
J^3_0 (m,h) & =& m (m,h) \\
L_0 (m,h) & = & h (m,h) \,.
\end{eqnarray}
On the other hand, the diagonal array of circles represents the
states $(m-\onth,m-\onth)$, $m\in\Zop$. The two arrays intersect at
$(-\onth,-\onth)$, and the circle at this intersection corresponds to
the state $(-\onth,-\onth)_1$. The arrows indicate the action of
$J^\pm_0$ (for the horizontal line) and $J^\pm_{\mp 1}$ (for the
diagonal line).  Finally, the empty circle represents the state
$(-\onth,-\onth)_2$ whose position in the charge lattice has been
slightly shifted so that it does not lie on top of the other state
with these charges. More specifically, we have the relations 
\begin{eqnarray}
J^+_0 \left(-\onth,-\onth\right) & = & \left(\twth,-\onth\right) \,, \\
J^-_0 \left(\twth,-\onth\right) & = & 
J^+_{-1} \left(-{4\over 3},-{4\over 3} \right)\,, 
\end{eqnarray}
and
\be
J^-_0 J^+_0 \left(-\onth,-\onth\right)_1 = 
{1\over 3} \left(-\onth,-\onth\right)_2 = 
J^+_{-1} J^-_{1} \left(-\onth,-\onth\right)_1 \,.
\ee
Furthermore we find that 
\begin{eqnarray}
L_0 \left(-\onth,-\onth\right)_1 & = &
- \onth \left(-\onth,-\onth\right)_1 + \left(-\onth,-\onth\right)_2
\\
L_0 \left(-\onth,-\onth\right)_2 & = &
- \onth \left(-\onth,-\onth\right)_2\,.
\end{eqnarray}
Thus the representation $\R_{-\onth}$ is in fact a logarithmic
representation. This does not contradict what we have found before
since $\R_{-\onth}$ is not a highest weight representation. In fact,
it was shown in \cite{Gab01} that with these relations one finds that 
\be
V_0(\Nu) \left(-\onth,-\onth\right)_1 = 0 \,,
\ee
where $\Nu$ is the null-vector in (\ref{vacnullsu2}).
\bigskip

\noindent The representation $\R_{\onth}$ is not the only logarithmic
representation that occurs in fusion products of the Kac-Wakimoto
representations. In fact, the relevant representations also contain an 
extension of the vacuum representation $\R_0$
\be
\left(E_0 \otimes E_0 \right)_{\mathrm f} = \R_0 \oplus E_0\,,
\ee
whose structure can be schematically described by figure~2.
\begin{figure}[htb]
\epsfysize=6cm
\centerline{\epsffile{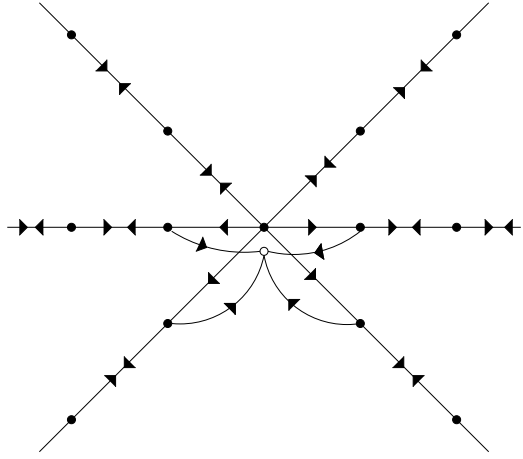}}
\caption{The representation $\R_0$.}
\end{figure}

\noindent Here, as before, the different states have been arranged
according to their $J^3_0$ and $L_0$ charge. The horizontal line
represents states with charges $(m,0)$, $m\in\Zop$, and the arrows
correspond to the action of $J^\pm_0$. The state $(0,0)$ corresponds
to $\omega$, and  the empty circle represents the state $\Omega$ whose
position in the charge  lattice has been slightly shifted so that it
does not lie on top of the state $\omega$. The arrows along the
diagonal arrays correspond to the action of $J^\pm_{\pm 1}$ and
$J^\pm_{\mp 1}$, respectively. More specifically, we have the
relations  
\be
J^-_{-1} J^+_{1} \omega  = \Omega = J^+_{-1} J^-_{1} \omega \,,
\ee
\be
J^-_0 J^+_0 \omega  = \gamma \Omega = J^-_0 J^+_0 \omega \,,
\ee
and
\be
L_0 \omega  = \left(3+{3\over 2} \gamma\right)\Omega\,,
\ee
where $\gamma\ne 0,-2$ is a constant (that was not determined in
\cite{Gab01}). Thus the two states $\omega$ and $\Omega$ form again
the familiar Jordan block!
\bigskip

One can actually determine at least some of the remaining fusion rules
explicitly; under some natural assumption (that is described in
\cite{Gab01}) one then finds that the fusion rules close among the
representations $\H_0$, $D^\pm_j$, $E_0$, $\R_0$ and $\R_{\onth}$,
together with their images under $\pi_s$. More specifically, one finds 
\begin{eqnarray}
\left( D^-_{\onth} \otimes \R_0 \right)_{\mathrm f}  
           & = & \R_{\onth}\\
\left( D^-_{\onth} \otimes \R_{\onth} \right)_{\mathrm f}  
           & = & 2\, \pi_1 (E_0) \oplus \pi_1 (\R_0) \\
\left( E_0 \otimes \R_0 \right)_{\mathrm f}  
           & = & 2\, E_0 \\
\left( E_0 \otimes \R_{\onth} \right)_{\mathrm f}  
           & = & 2\, \R_{\onth} \\
\left( \R_0 \otimes \R_0 \right)_{\mathrm f}  
           & = & 2\, \R_0 \\
\left( \R_0 \otimes \R_{\onth} \right)_{\mathrm f}  
           & = & 2\, \R_{\onth} \\
\left( \R_{\onth} \otimes \R_{\onth} \right)_{\mathrm f}  
           & = & 2\, \pi_1(\R_0) \oplus 4\, \pi_1 (E_0)\,.
\end{eqnarray}
Together with (\ref{symmetry}) this then determines all fusion
products. 

\newpage

\section{Conclusions}
\renewcommand{\theequation}{6.\arabic{equation}}
\setcounter{equation}{0}

In these lectures we have described how logarithmic conformal field
theories can be understood from an algebraic (representation
theoretic) point of view. We have explained how this point of view is
related to the description of logarithmic conformal field theory in
terms of amplitudes with logarithmic branch cuts. We have also shown 
that the algebraic approach is quite general and that it allows one to
describe logarithmic representations that would be difficult to
characterise otherwise. In particular, this is the case for
logarithmic representations that are not obtained by the action of the
chiral algebra from a highest weight state.  

We have mainly considered the triplet theory at $c=-2$ which is the
best understood (rational) logarithmic conformal field theory. We have
analysed its fusion rules in detail, and we have described how to
construct the corresponding local theory. Most of the logarithmic
conformal field theories that have been studied so far are closely
related to the $c=-2$ Virasoro model or an extension thereof. 

In the last section we have shown that the $su(2)$ model at $k=-4/3$
is also a logarithmic conformal field theory. This result suggests
that the same will presumably be the case for all other WZW models at
fractional level. The logarithmic representations of these theories
are presumably not highest weight representations, and their structure
is therefore likely to be subtle. It would be very interesting to
study these representations in more detail, and to get some better
understanding of their general structure.
\smallskip

Given that already WZW models at fractional level appear to be
logarithmic, it seems reasonable to speculate that most non-rational
theories will actually be logarithmic. If one wants to obtain some
more general understanding of non-rational theories, it will in
particular be necessary to get to grips with logarithmic conformal
field theories. It is hoped that the present notes will prove useful
for that purpose.

\newpage

\section*{Acknowledgements}

I would like to thank the organisers for organising an interesting and
very stimulating school, and for giving me the opportunity to present
these lectures. I am in particular grateful to Shahin Rouhani and Reza
Rahimi Tabar for their wonderful hospitality. 

I thank Horst Kausch and Peter Goddard for many discussions and
explanations over the years. I am also grateful to Michael Flohr and
Shahin Rouhani for many useful conversations during the school, as
well as to the audience for asking good questions.

I am grateful to the Royal Society for a University Research    
Fellowship.

\newpage

\appendix

\section*{Appendix A: The algebra structure of Zhu's algebra}
\renewcommand{\theequation}{A.\arabic{equation}}
\setcounter{equation}{0}

In this appendix we want to show that the product defined by
(\ref{prodd}) makes $A(\H_0)$ into an algebra. In order to exhibit the
structure of this algebra it is useful to introduce a second set of
modes by   
\be
\label{zhumodec}
V^{(N)}_c(\psi) 
= (-1)^N \oint {d w \over w} {1 \over (w+1)} 
\left({w+1 \over w}\right)^N V\left( (w+1)^{L_0} \psi, w \right)\,.
\ee
These modes are characterised by the property that 
\begin{eqnarray}
\langle \bar\phi | \phi(-1) \;V^{(N)}_c(\psi) \; \chi\rangle & = & 0
\qquad \hbox{for $N>0$,} \\
\langle \bar\phi | \phi(-1) \;V^{(0)}_c(\psi) \; \chi\rangle & = &
\langle \bar\phi | \left( V_0(\psi) \phi\right)(-1) \; \chi\rangle\,.
\label{zhuaction1}
\end{eqnarray}
It is obvious from (\ref{zhumode}) and (\ref{zhumodec}) that 
$$
V^{(1)}(\psi)=V^{(1)}_c(\psi)\,,
$$ 
and the analogue of (\ref{recursive}) is 
\be
\label{recursive1}
V^{(N+1)}_c(\psi) = - {1 \over N+1} V^{(N)}_c(L_{-1}\psi)
-  {N+h_\psi\over N+1} V^{(N)}_c(\psi) \,.
\ee
The space $O(\H_0)$ is therefore also generated by the states 
$$
O(\H_0) = \hbox{span} 
\left\{V^{(1)}_c(\psi)\chi \, : \, \psi,\chi\in\H_0\right\}\,.
$$
Let us introduce, following (\ref{zhuaction}) and (\ref{zhuaction1}),
the notation    
\begin{eqnarray}
V_L(\psi) & \equiv & V^{(0)}(\psi) \nonumber \\
V_R(\psi) & \equiv & V^{(0)}_c(\psi) \nonumber \\
N(\psi) & = & V^{(1)}(\psi)=V^{(1)}_c(\psi) \,. \nonumber
\end{eqnarray}
We also denote by $N(\H_0)$ the vector space of operators that are
spanned by $N(\psi)$ for $\psi\in\H_0$; then $O(\H_0)=N(\H_0) \H_0$. 
Finally it follows from the fact that the vacuum $\Omega$ is
annihilated by all modes $V_n(\psi)$ with $n>-h_\psi$ together with 
$V_{-h_\psi}(\psi)\Omega=\psi$ that 
\be
V_L(\psi)\Omega =V_R(\psi)\Omega=\psi\,.
\ee

The equations (\ref{zhuaction}) and (\ref{zhuaction1}) suggest that
the modes $V_L(\psi)$ and $V_R(\chi)$ commute up to an operator in
$N(\H_0)$. In order to prove this it is sufficient to consider the
case where $\psi$ and $\chi$ are both eigenvectors of $L_0$ with
eigenvalues $h_\psi$ and $h_\chi$, respectively. Then we have 
\begin{eqnarray}
\label{commutator}
{}[V_L(\psi),V_R(\chi)] & = &
\oint\oint_{|\zeta|>|w|} {d\zeta \over \zeta}(\zeta+1)^{h_\psi}
{dw \over w } (w+1)^{h_\chi-1}
V(\psi,\zeta) V(\chi,w) \nonumber \\
& & \qquad 
- \oint\oint_{|w|>|\zeta|} {dw \over w} 
(w+1)^{h_\chi -1} {d\zeta \over \zeta} 
(\zeta+1)^{h_\psi} V(\chi,w) V(\psi,\zeta) \nonumber \\
& = & 
\oint_0 \left\{ \oint_w {d\zeta \over \zeta}
(\zeta+1)^{h_\psi}  V(\psi,\zeta) V(\chi,w) \right\} 
{dw \over w} (w+1)^{h_\chi-1}  \nonumber \\
& = & \sum_{n}
\oint_0 \left\{ \oint_w {d\zeta \over \zeta}
(\zeta+1)^{h_\psi} V(V_n(\psi)\chi,w) (\zeta-w)^{-n-h_\psi} \right\}
\nonumber \\
& & \qquad \qquad 
{dw \over w} (w+1)^{h_\chi-1} \nonumber \\
& = &
\sum_{h_\chi\geq n \geq 0} \sum_{l=0}^{n+h_\psi - 1}
(-1)^l \, {h_\psi \choose l+1-n} \nonumber \\
&  &  \qquad \qquad 
\oint_0 {dw \over w (w+1)} \left( {w+1 \over w}\right)^{l+1}
(w+1)^{h_\chi-n} V(V_n(\psi)\chi,w) \nonumber \\
& \in & N(\H_0)\,.
\end{eqnarray}
Because of (\ref{recursive1}), every element in $N(\H_0)$ can be written
as $V_R(\phi)$ for a suitable $\phi$, and (\ref{commutator}) thus
implies that $[V_L(\psi),N(\chi)] \in N(\H_0)$; hence
$V_L(\psi)$ defines an endomorphism of $\A(\H_0)$. 

For two endomorphisms, $\Phi_1, \Phi_2$, of $\H_0$, which leave
$O(\H_0)$ invariant (so that they induce endomorphisms of $\A(\H_0))$, 
we shall write $\Phi_1\approx\Phi_2$ if they agree as endomorphisms of 
$\A(\H_0)$, \ie\ if $(\Phi_1-\Phi_2)\H_0\subset O(\H_0)$. Similarly 
we write $\phi\approx 0$ if $\phi\in O(\H_0)$.

In the same way in which the action of $V(\psi,z)$ is uniquely
characterised by locality and its action on the vacuum (see
\cite{God}), we can now prove the following  

\noindent {\bf Uniqueness Theorem for Zhu modes} \cite{GabGod}:
Suppose $\Phi$ is an endomorphism of $\H_0$ that leaves $O(\H_0)$
invariant and satisfies  
\begin{eqnarray}
\Phi\Omega & =& \psi \nonumber \\
{}[\Phi,V_R(\chi)] & \in & N(\H_0) \quad 
\hbox{for all $\chi\in\H_0$.} \nonumber 
\end{eqnarray}
Then $\Phi\approx V_L(\psi)$. 

\noindent {\bf Proof}: This follows from 
$$
\Phi \, \chi=\Phi \, V_R(\chi)\Omega \approx V_R(\chi)\, \Phi\, \Omega=
V_R(\chi) \psi = V_R(\chi) \, V_L(\psi) \, \Omega \approx 
V_L(\psi)\chi \,,
$$
where we have used that $V_L(\psi)\Omega=V_R(\psi)\Omega=\psi$.
\medskip

\noindent It is then an immediate consequence that
\be
\label{A}
V_L(V_L(\psi)\chi)\approx V_L(\psi)V_L(\chi) \,,
\ee
and a particular case of this (using again the fact that every element
in $N(\H_0)$ can be written as $V_L(\phi)$ for some suitable $\phi$) 
is that 
\be
V_L(N(\psi)\chi)\approx N(\psi) V_L(\chi)\,. 
\ee
In particular this implies that the product (\ref{prodd})
$\phi\ast_L\psi$  is well-defined for both
$\phi,\psi\in\A(\H_0)$. Furthermore, (\ref{A}) shows that this product
is associative, and thus $\A(\H_0)$ has the structure of an algebra. 

\noindent We can also define a product by
$\phi\ast_R\psi=V_R(\phi)\psi$. Since  
$$
\phi\ast_L\psi= V_L(\phi)\psi= V_L(\phi)V_R(\psi)\Omega\approx
V_R(\psi)V_L(\phi)\Omega=V_R(\psi)\phi=\psi\ast_R\phi
$$
this defines the reverse ring (or algebra) structure.

\end{document}